\documentclass[
aps,
prl,
amsmath, amssymb,
twocolumn,
superscriptaddress,
reprint,
nofootinbib,
floatfix,
noeprint
]{revtex4-2}

\usepackage[english]{babel}
\usepackage[utf8]{inputenc}
\usepackage[T1]{fontenc}

\usepackage{bbm}
\usepackage{physics}
\usepackage{subfigure}
\usepackage{mathrsfs}
\usepackage{bm}
\usepackage{enumerate}
\usepackage{graphicx}
\usepackage[dvipsnames]{xcolor}
\usepackage{float}
\usepackage{subdepth}
\usepackage{booktabs}
\usepackage{multirow}
\usepackage[math]{cellspace}

\usepackage[colorlinks,
linkcolor=BrickRed,
citecolor=MidnightBlue,
urlcolor=MidnightBlue,
bookmarks=true,
bookmarksopen=true,
bookmarksnumbered=true,
]{hyperref}

\newcommand{\av}[1]{\left\langle#1\right\rangle}

\renewcommand{\i}{\mathrm{i}}
\renewcommand{\(}{\left(}
\renewcommand{\)}{\right)}
\newcommand{\id}{\mathbbm{1}}

\renewcommand{\mod}{\mathrm{\,mod\,}}

\newcommand{\be}{\begin{eqnarray}}
	\newcommand{\bea}{\begin{eqnarray}}
		\newcommand{\eea}{\end{eqnarray}}
	\newcommand{\beq}{\begin{equation}}
		\newcommand{\ee}{\end{eqnarray}}
	\newcommand{\eeq}{\end{equation}}

\newcounter{ls}

\newcounter{jvcc}
\newcounter{amgg}
\newcounter{yc}

\begin{document}
\title{Sixfold Way of Traversable Wormholes in the Sachdev-Ye-Kitaev Model}

\author{Antonio M. Garc\'\i a-Garc\'\i a}
\email{amgg@sjtu.edu.cn}
\affiliation{Shanghai Center for Complex Physics,
	School of Physics and Astronomy, Shanghai Jiao Tong
	University, Shanghai 200240, China}

\author{Lucas S\'a}
\email{lucas.seara.sa@tecnico.ulisboa.pt}
\affiliation{TCM Group, Cavendish Laboratory, University of Cambridge, JJ Thomson Avenue, Cambridge CB3 0HE, UK\looseness=-1}
\affiliation{CeFEMA, Instituto Superior T\'ecnico, Universidade de Lisboa, Av.\ Rovisco Pais, 1049-001 Lisboa, Portugal}

\author{Jacobus J. M. Verbaarschot}
\email{jacobus.verbaarschot@stonybrook.edu}
\affiliation{Center for Nuclear Theory and Department of Physics and Astronomy, Stony Brook University, Stony Brook, New York 11794, USA}

\author{Can Yin
%	\begin{CJK*}{UTF8}{gbsn}
%		(殷灿)
%\end{CJK*}
}
\email{yin_can@sjtu.edu.cn}
\affiliation{Shanghai Center for Complex Physics,
	School of Physics and Astronomy, Shanghai Jiao Tong
	University, Shanghai 200240, China}

\begin{abstract}
  In the infrared limit, a nearly anti-de Sitter spacetime in two dimensions (AdS$_2$) perturbed by a weak double trace deformation and a two-site $(q>2)$-body Sachdev-Ye-Kitaev (SYK) model with $N$ Majoranas and a weak $2r$-body intersite coupling share the same near-conformal dynamics described by a traversable wormhole.
  We exploit this relation to propose a symmetry classification of traversable wormholes depending on $N$, $q$, and $r$, with $q>2r$, and confirm it by a level statistics analysis using exact diagonalization techniques. Intriguingly, a time-reversed state never results in a new state, so only six universality classes occur---A, AI, BDI, CI, C, and D---and different symmetry sectors of the model may belong to distinct universality classes.
\end{abstract}

\maketitle

A generic many-body quantum chaotic system that does not suffer from localization~\cite{anderson1958,basko2006} eventually reaches an ergodic state governed by the symmetries of the system, rather than by the microscopic details of its Hamiltonian. Since this ergodic state only depends on global symmetries, it is possible to classify the dynamics by these symmetries. The study of level statistics is a powerful tool for establishing this classification because the level statistics of quantum chaotic systems~\cite{bohigas1984,guhr1998} agree with the predictions of random matrix theory (RMT)~\cite{wigner1951,dyson1962a,dyson1962b,dyson1962c,dyson1962d,dyson1972,mehta2004}.
Based on the theory of symmetric spaces, it was concluded that, after taking care of unitary symmetries, only ten universality classes exist, the so-called tenfold way of RMT~\cite{altland1997}.
The symmetry classification was later extended to non-Hermitian systems, where 38 universality classes exist~\cite{bernard2002,kawabata2019,zhou2019,hamazaki2020,kanazawa2021,garcia2022d}.

The ten universality classes of Hermitian quantum chaotic systems have already been identified. Three classes related to the presence of time-reversal symmetry (TRS) [an antiunitary operator that commutes with the Hamiltonian] were reported in early studies in nuclear physics~\cite{wigner1951,dyson1962a} and single-particle quantum chaotic systems~\cite{bohigas1984,efetov1983}, pertaining to systems with time-reversal invariance (class AI), broken time-reversal invariance (class A), and time-reversal invariance with broken rotational invariance and half-integer spin (class AII).
Ensembles of antisymmetric~\cite{mehta1968distribution} and anti-self-dual~\cite{mehta1983some} Hermitian random matrices (classes D and C, respectively) were also discovered early on.
Later, studies of the spectral properties of the
QCD Dirac operator~\cite{shuryak1993a,verbaarschot1993a} revealed the existence of three more universality classes related to chiral symmetries represented by a unitary operator that anticommutes with the Hamiltonian (classes AIII, BDI, and CII).
Shortly afterward, the classification was completed by adding chiral matrices with symmetric and antisymmetric off-diagonal blocks (classes CI and DIII, respectively)~\cite{altland1997}. Physically, these classes are realized in superconducting systems with particle-hole symmetry (PHS) [an antiunitary operator that anticommutes with the Hamiltonian].

A related question is how many of the universality classes can be identified in more specific Hamiltonians describing a certain phenomenon. For instance, a full classificatory scheme was worked out for topological insulators in Ref.~\cite{ryu2010} and for systems at the Anderson transition~\cite{anderson1958} in Refs.~\cite{garcia2000,garcia2002,garcia2003,evers2008}.
More recently, the Sachdev-Ye-Kitaev (SYK) model~\cite{french1970,bohigas1971,bohigas1971a,french1971,mon1975,verbaarschot1984,benet2001,sachdev1993,kitaev2015,borgonovi2016,maldacena2016} has been classified~\cite{you2016,garcia2016,Cotler:2016fpe,altland:2017eao,li2017,kanazawa2017,beri2019,beri2020,sun2020,garcia2021d} in terms of RMT, thus
providing a symmetry classification of quantum black holes in two-dimensional nearly anti-de Sitter (AdS$_2$) backgrounds.
The relation between quantum gravity and the SYK model~\cite{kitaev2015,maldacena2016,bagrets2016,Altland:2022xqx,Berkooz:2023cqc}
has been extended to traversable~\cite{maldacena2018},
Euclidean~\cite{maldacena2018,garcia2021}, and Keldysh~\cite{garcia2022e} wormholes.
A study of level statistics~\cite{garcia2019} revealed that traversable wormholes~\cite{maldacena2018} belong to the universality class of systems with TRS (class AI).
A natural question to ask is whether this symmetry is a necessary condition for the existence of traversable wormholes. %or other symmetry classes are allowed.

The main goal of this Letter is to answer this question by providing an explicit symmetry classification of SYK configurations whose gravity dual is a nearly AdS$_2$ traversable wormhole that encompasses six universality classes. For that purpose, we introduce a two-site, left ($L$) and right ($R$), Hermitian SYK Hamiltonian $H=H_L+\alpha (-1)^{q/2} H_R +\lambda H_I$ ~\cite{maldacena2018,Klebanov:2020kck,Kim:2019upg,Arefeva:2019ugp,Maldacena:2019ufo}
with left-right asymmetry parameter $\alpha$ and coupling constant $\lambda$. The two single-site $q$-body SYK Hamiltonians $H_{L,R}$ of $N$ Majorana fermions
and the $2r$-body Hamiltonian $H_I$ coupling them
are given by
\be\label{eq:hamdef}
  H_{L,R}&=&\i^{q/2} \sum_{i_1<\cdots<i_q}^{N} J_{i_1\cdots i_q} \, \psi^{L,R}_{i_1} \cdots \, \psi^{L,R}_{i_q},\\
H_I&=&\i^r \frac{N^{1-r}}{r} \left(\sum_{i=1}^{N}\psi^L_{i}\psi^R_{i}\right)^r,
\ee
where the couplings $J_{i_1\cdots i_q}$ are Gaussian random variables with zero mean and variance $\sigma^2=2^{q-1}(q-1)!/(q N^{q-1})$.
The Majorana fermions satisfy the commutation relation $\{ \psi_{i}^A, \psi_{j}^B \} =\delta_{AB}\delta_{ij}$ ($i,j=1,\dots, N$ and $A,B=L,R$). The parameters
$N$ and $q$ are taken to be even [see the Supplemental Material (SM)~\cite{SM} for odd $N$].

\nocite{fu2018,Kieburg:2014eca,garcia2018a}

{\it Symmetry classification.---}%
The class of $H$ is determined by its behavior under antiunitary symmetries:\\
1. TRS: $THT^{-1}=+H$, $T\i T^{-1}=-\i$, $T^2=\pm1$;\\
2. PHS: $CHC^{-1}=-H$, $C\i C^{-1}=-\i$, $C^2=\pm1$.\\
Any antiunitary symmetry $A$ can be decomposed as $A=UK$, with $U$ unitary and $K$ complex conjugation, and we choose a basis such that
\begin{align}
\label{eq:main_action_K_psi}
K \psi_i^L K^{-1} = \psi_i^L,
\qquad
K \psi_i^R K^{-1} = -\psi_i^R.
\end{align}

We define the unitary left and right parities,
\begin{equation}
S_L=(2\i)^{N/2}\prod_{i=1}^N\psi_i^L,
\qquad
S_R=(2\i)^{N/2}\prod_{i=1}^N\psi_i^R,
\end{equation}
the total parity, $S=S_LS_R$, and the (exponential of the) spin operator,
\begin{equation}
Q=\exp\left\{-\frac{\pi}{4}\sum_{i=1}^N\psi_i^L\psi_i^R\right\}
=2^{-N/2}\prod_{i=1}^N\(1-2\psi_i^L\psi_i^R\).
\end{equation}
They square to $S_L^2=S_R^2=S^2=+1$ and $Q^2=S$. Their action on the Majorana fermions is given by
(recall that, throughout the main text, $N$ is even)
\begin{align}
\nonumber
\label{eq:main_action_SLRQ_psi}
&S_{L,R} \psi_i^{L,R} S_{L,R}^{-1} = -\psi_i^{L,R},
\quad
&&S_{L,R} \psi_i^{R,L} S_{L,R}^{-1} = \psi_i^{R,L},
\\
&Q \psi_i^R Q^{-1} = -\psi_i^L,
\quad
&&Q \psi_i^L Q^{-1} = \psi_i^R,
\end{align}
i.e., the parity operators act inside each site, flipping the sign of the respective Majoranas, while, up to a sign, the spin operator exchanges the two species. Other operations are obtained by compositions, e.g., the total parity $S$ reverses the signs of the Majoranas in both sites simultaneously.
Using Eqs.~(\ref{eq:main_action_K_psi}) to (\ref{eq:main_action_SLRQ_psi}), the transformation properties of $H_0\equiv H_L + \alpha (-1)^{q/2} H_R$ and $H_I$ under the unitary and antiunitary operators are:
\begin{align}
\label{eq:main_action_SLRQ_H}
\nonumber
&S_{L,R} H_0 S_{L,R}^{-1} =H_0,
&&S_{L,R} H_I S_{L,R}^{-1} = (-1)^r H_I,
\\
\nonumber
&Q H_0 Q^{-1} = (-1)^{q/2} H_0,
&&Q H_I Q^{-1} = H_I,
\\
&K H_0 K^{-1} = (-1)^{q/2} H_0,
&&K H_I K^{-1} = H_I.
\end{align}
Note the transformation of $H_0$ under $Q$ holds only for $\alpha=1$. If $\alpha\neq1$, the left-right spin operator $Q$ is not a symmetry of $H$.

The symmetry classification of $H$, which follows from Eq.~(\ref{eq:main_action_SLRQ_H}), depends on $N\mod4$~\cite{kawabata2022} (in contrast to the one-site classification which depends on $N\mod8$~\cite{you2016,garcia2016,Cotler:2016fpe,beri2019}), the parity of $q/2$ and $r$, and whether there is left-right symmetry ($\alpha=1$) or not ($\alpha\neq1$), see Tables~\ref{tab:classification_even_even_even}--\ref{tab:classification_even_odd_odd}. Below we provide a brief justification of this classification, see the SM~\cite{SM} for a detailed derivation.
\begin{table}[t]
	\caption{Symmetry classification of the two-site SYK Hamiltonian for even $q/2$ and even $r$. Each line corresponds to a block of the Hamiltonian, labeled by the eigenvalues of the conserved quantities $S_{L,R}$ and $Q$. For each of the six blocks, we give its dimension and its symmetry class both for the left-right symmetric ($\alpha=1$) and asymmetric ($\alpha\neq 1$) cases and for $N\mod4=0,2$.}
	\label{tab:classification_even_even_even}
	\begin{tabular}{@{}Sc Sc Sc Sc Sc Sc Sc Sc@{}}
		\toprule
		\multirow{2}{*}{$S_L$} & \multirow{2}{*}{$S_R$} & \multirow{2}{*}{$Q$} & \multirow{2}{*}{Dimension} & \multicolumn{2}{c}{$N\mod4=0$}   & \multicolumn{2}{c}{$N\mod4=2$}    \\
		&                        &                      &                       & $\alpha=1$ & $\alpha\neq1$       & $\alpha=1$ & $\alpha\neq1$       \\ \midrule
		\multirow{2}{*}{$+1$}  & \multirow{2}{*}{$+1$}  & $+1$                &  $(2^N+2^{N/2+1})/8$                     & AI         & \multirow{2}{*}{AI} & A          & \multirow{2}{*}{A} \\
		&                        & $-1$                & $(2^N-2^{N/2+1})/8$                      & AI         &                     & A          &                     \\ \midrule
		\multirow{2}{*}{$-1$}  & \multirow{2}{*}{$-1$}  & $+1$                &  $(2^N+2^{N/2+1})/8$  & AI         & \multirow{2}{*}{AI} & A          & \multirow{2}{*}{A} \\
		&                        & $-1$                &  $(2^N-2^{N/2+1})/8$                     & AI         &                     & A          &                     \\ \midrule
		$+1$                   & $-1$                   & ---                  &  $2^N/4$                     & AI         & AI                  & AI         & A                  \\ \midrule
		$-1$                   & $+1$                   & ---                  &  $2^N/4$                     & AI         & AI                  & AI         & A                  \\ \bottomrule
	\end{tabular}
	
	\caption{Same as Table~\ref{tab:classification_even_even_even}, but for even $q/2$ and odd $r$. There are four blocks labeled by the eigenvalues of $S$ and $Q$. The results are independent of $N$.}
	\label{tab:classification_even_even_odd}
	\begin{tabular}{@{}Sc Sc Sc Sc Sc@{}}
		\toprule
		$S$  & $Q$   & Dimension                & $\alpha=1$ & $\alpha\neq1$         \\ \midrule
		\multirow{2}{*}{$+1$} & $+1$  & $(2^N+2^{N/2+1})/4$ & AI         & \multirow{2}{*}{AI}   \\
		                      & $-1$  & $(2^N-2^{N/2+1})/4$ & AI         &                       \\ \midrule
		\multirow{2}{*}{$-1$} & $+\i$ & $2^N/4$             & AI         & \multirow{2}{*}{AI}   \\
		                      & $-\i$ & $2^N/4$             & AI         &                       \\ \bottomrule
	\end{tabular}
	
	\caption{Same as Table~\ref{tab:classification_even_even_even}, but for odd $q/2$ and even $r$. There are four blocks labeled by the eigenvalues of $S_{L,R}$.}
	\label{tab:classification_even_odd_even}
	\begin{tabular}{@{}Sc Sc Sc Sc Sc Sc Sc@{}}
		\toprule
		\multirow{2}{*}{$S_L$} & \multirow{2}{*}{$S_R$} & \multirow{2}{*}{Dimension} & \multicolumn{2}{c}{$N\mod4=0$}   & \multicolumn{2}{c}{$N\mod4=2$}    \\
		&                        &   & $\alpha=1$   & $\alpha\neq1$  & $\alpha=1$ &  $\alpha\neq1$       \\ \midrule
		$+1$ & $+1$ & $2^N/4$ & AI & A & A  & A  \\
		$-1$ & $-1$ & $2^N/4$ & AI & A & A  & A  \\
		$+1$ & $-1$ & $2^N/4$ & A  & A & AI & A  \\
		$-1$ & $+1$ & $2^N/4$ & A  & A & AI & A  \\ \bottomrule
	\end{tabular}
	
	\caption{Same as Table~\ref{tab:classification_even_even_even}, but for odd $q/2$ and odd $r$. There are two blocks labeled by the eigenvalues of $S$.}
	\label{tab:classification_even_odd_odd}
	\begin{tabular}{@{}Sc Sc Sc Sc Sc Sc@{}}
		\toprule
		\multirow{2}{*}{$S$} & \multirow{2}{*}{Dimension} & \multicolumn{2}{c}{$N\mod4=0$}   & \multicolumn{2}{c}{$N\mod4=2$}  \\
		&                       & $\alpha=1$   & $\alpha\neq1$     & $\alpha=1$ &  $\alpha\neq1$     \\ \midrule
		$+1$ & $2^N/2$ & BDI  & D & CI & C  \\
		$-1$ & $2^N/2$ & BDI  & D & CI & C  \\ \bottomrule
	\end{tabular}
\end{table}

{\it First}, for the action of an antiunitary symmetry to be well-defined, all commuting unitary symmetries (conserved quantities) must be resolved. That is, in the common eigenbasis of $H$ and its unitary symmetries, $H$ assumes a block-diagonal structure, and the antiunitary symmetries $T$ and $C$ must act within a single block. The two-site SYK Hamiltonian can have two, four, or six blocks, as indicated in Tables~\ref{tab:classification_even_even_even}--\ref{tab:classification_even_odd_odd}. The total parity $S$ is always conserved and, thus, $H$ has at least two blocks identified by its eigenvalues $s=\pm1$. For $\alpha=1$, we have the following possibilities:\\
1. If $q/2$ and $r$ are both odd, there is no other unitary symmetry and there are only two blocks (Table~\ref{tab:classification_even_odd_odd}).\\
2. If $r$ is odd but $q/2$ is even, $Q$ is a symmetry of $H$ and, because of $QS=SQ$, it splits the two blocks into two subblocks each and we get four blocks, labeled by the eigenvalues of $S$, $s=\pm1$, and $Q$, $k=\pm1,\pm\i$ (Table~\ref{tab:classification_even_even_odd}).\\
3. When $r$ is even, $S_L$ and $S_R$ are independently conserved, defining, at least, four blocks, $s_{L,R}=\pm1$. If, moreover, $q/2$ is odd, these four blocks are the only blocks (Table~\ref{tab:classification_even_odd_even}).\\
4. If, instead, $q/2$ is even (with $r$ still even), and because $QS_{L,R}=S S_{L,R}Q$ [see Eq.~(\ref{eq:comm_QSR}) of the SM~\cite{SM}], the two blocks with $s=+1$ get split by $Q$ into two subblocks each, while the two blocks with $s=-1$ do not; in total, we get six blocks (Table~\ref{tab:classification_even_even_even}).
\\
If $\alpha\neq1$, the blocks of $S_{L,R}$ and $S$ are not split by $Q$ (four blocks for $r$ even, two for $r$ odd).

{\it Second}, TRS is implemented by either $T=K$ (for even $q/2$ and any $\alpha$) or $T=QK$ (for odd $q/2$ and $\alpha=1$). In either case, we have $T^2=+1$ and we conclude that $H$ displays the same level statistics as random matrices from either the Gaussian Orthogonal Ensemble (GOE) \cite{mehta2004,guhr1998}---if $T$ acts within a single block of $H$---, or the Gaussian Unitary Ensemble (GUE) \cite{mehta2004,guhr1998}---if $T$ connects different blocks. This is determined by the commutator of $T$ with the orthogonal projector onto the respective block, which must be checked on a case-by-case basis, see the SM~\cite{SM}. For odd $q/2$ and $\alpha\neq 1$, since $Q$ is not a symmetry of $H$, there is no TRS and all blocks display GUE statistics.

{\it Third}, for all cases except for $q/2$ and $r$ both odd, there is no PHS, then $T$ is the only antiunitary symmetry, and all blocks belong either to class AI (if $T$ acts within a single block) or class A (if it connects different blocks). When $q/2$ and $r$ are both odd, there exists a PHS implemented by $C=S_LK$, which squares to $C^2=(-1)^{N(N-1)/2}$, and commutes with the projector into a block with fixed $S$. In the left-right symmetric case ($\alpha = 1$), we thus have simultaneous TRS and PHS, and the blocks of $H$ belong to class BDI (for $N\mod4=0$) or CI (for $N\mod4=2$). In the asymmetric case ($\alpha\neq1$), we have only PHS and the blocks belong to class D ($N\mod4=0$) or C ($N\mod4=2$). Therefore, a slight asymmetry, $\alpha \approx 1$, substantially changes the universality class.

Comparing our results with the tenfold way~\cite{altland1997}, we can state the main results of this Letter. We have found a sixfold classification of the two-site SYK model Eq.~(\ref{eq:hamdef}): classes A, AI, BDI, CI, C, and D. Remarkably, for some parameters, different blocks of the same Hamiltonian belong to distinct symmetry classes, in contrast to the single-site SYK model~\cite{kanazawa2017}. Of the four remaining classes, class AIII---also not found in the standard single-site SYK model---could be realized by a Wishart extension of the model~\cite{garcia2021d} based on the product of two SYKs with complex-conjugated couplings. On the other hand, no classes with symplectic symmetry [i.e., $T^2=-1$, classes AII, CII, and DIII]---whose level statistics in the bulk are given by the Gaussian Symplectic Ensemble (GSE)---occur in the classification. The absence of these three classes, which indicates that a time-reversed state never results in a new state, arises as a fundamental restriction from the left-right symmetric intersite coupling of two SYKs. Universality classes with symplectic symmetry can still be realized if one considers a model with an asymmetric interaction, see the SM~\cite{SM}. The absence of AII$^\dagger$ statistics (the equivalent of GSE statistics in non-Hermitian systems~\cite{hamazaki2020}) has been observed recently in Lindbladian quantum dissipative dynamics~\cite{sa2022lindblad,kawabata2022}, which by construction has left-right symmetry. Therefore, generic coupled quantum systems with a left-right symmetric interaction do not have this symplectic symmetry.

{\it Level statistics.---}%
To confirm the proposed symmetry classification, we compare level correlations for different choices of parameters ($N$, $q$, $r$, and $\alpha$), with the predictions of RMT for the corresponding universality classes. This procedure is justified because the SYK model is quantum chaotic and deviations from RMT only affect a few eigenvalues close to the ground state~\cite{garcia2019}. The spectrum of the Hamiltonian (\ref{eq:hamdef}) is obtained by exact diagonalization techniques. At least $10^{5}$ eigenvalues are used for a given set of parameters.

\begin{figure}[t]
	\centering
	\includegraphics[width=7cm]{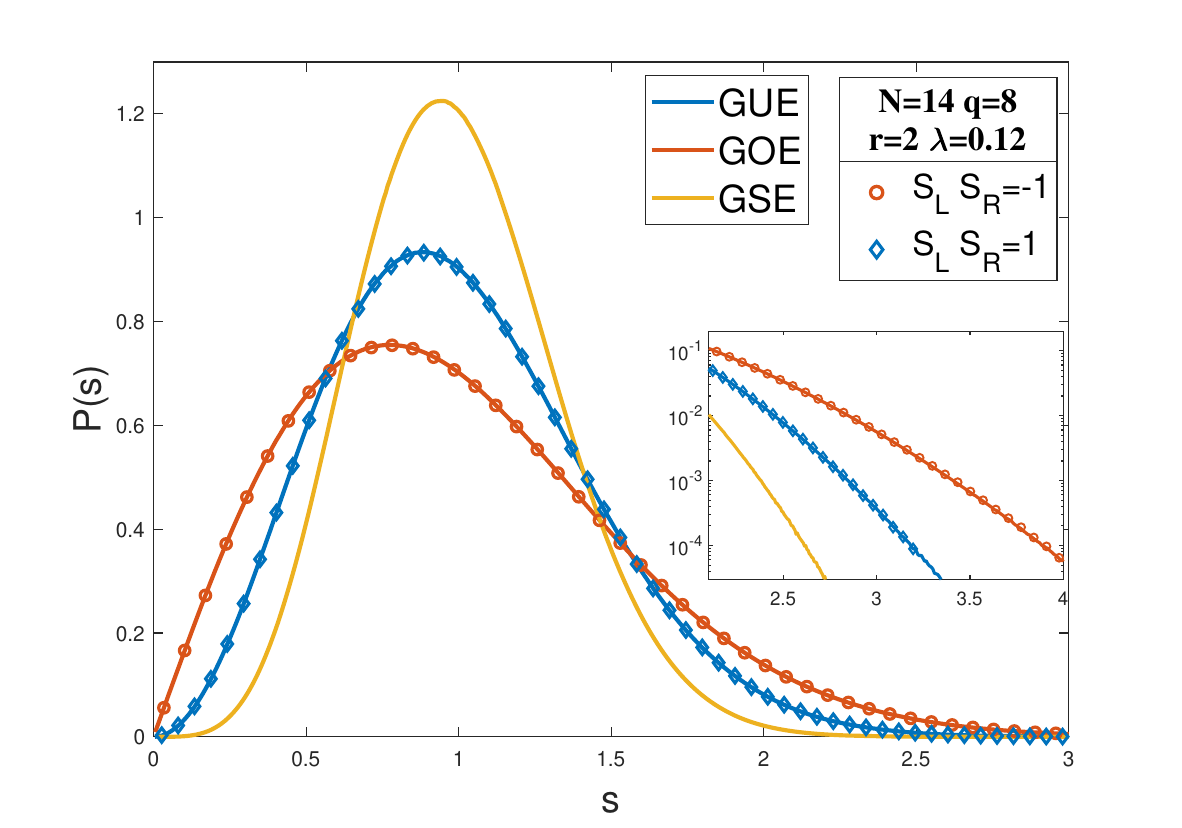}
	\caption{Level spacing distribution $P(s)$ for $N=14$, $r=2$, $q=8$, $\lambda =0.12$, and $\alpha = 1$
		that belong to the symmetry class A (GUE) or AI (GOE) depending on whether $S=S_LS_R=\pm1$, see Table~\ref{tab:classification_even_even_even}.
		We find excellent agreement with RMT even in the tail of the spectrum (inset).
	}\label{fig:Dyson}
\end{figure}

For the study of classes A and AI, we employ the distribution $P(s)$ of the level spacings $s_i = (E_i- E_{i-1})/\Delta$, where $E_i$ is the set of ordered eigenvalues and $\Delta$ is the mean level spacing. We unfold the spectrum~\cite{guhr1998} using a low-order (at most sixth) polynomial fitting. We have found that blocks with $T^2=+1$ (class AI) exhibit GOE level statistics, while blocks without TRS (class A) display GUE statistics. For $\lambda = 0$, both sites are uncorrelated, so the level statistics are given by Poisson statistics. Likewise, when $\lambda \rightarrow \infty$, the integrable $H_{I}$ dominates and thus level statistics are not given by RMT either. Therefore, it is necessary to choose an intermediate value of $\lambda$, so that levels are sufficiently mixed by the interaction. As an illustrative example, Fig.~\ref{fig:Dyson} depicts the level spacing distribution in a case with even $q/2$ and $r$, $N\mod4=2$, and $\alpha=1$, where the classification predicts class A (in blocks with total parity $s=+1$) or AI ($s=-1$). These results confirm the agreement with the RMT prediction for the expected universality class, even in the tail of the distribution. An exhaustive confirmation of all the remaining cases, employing the spacing ratio distribution~\cite{oganesyan2007,atas2016}, is presented in the SM~\cite{SM}.

\begin{figure}[t!]
	\centering
	\includegraphics[width=4.25cm]{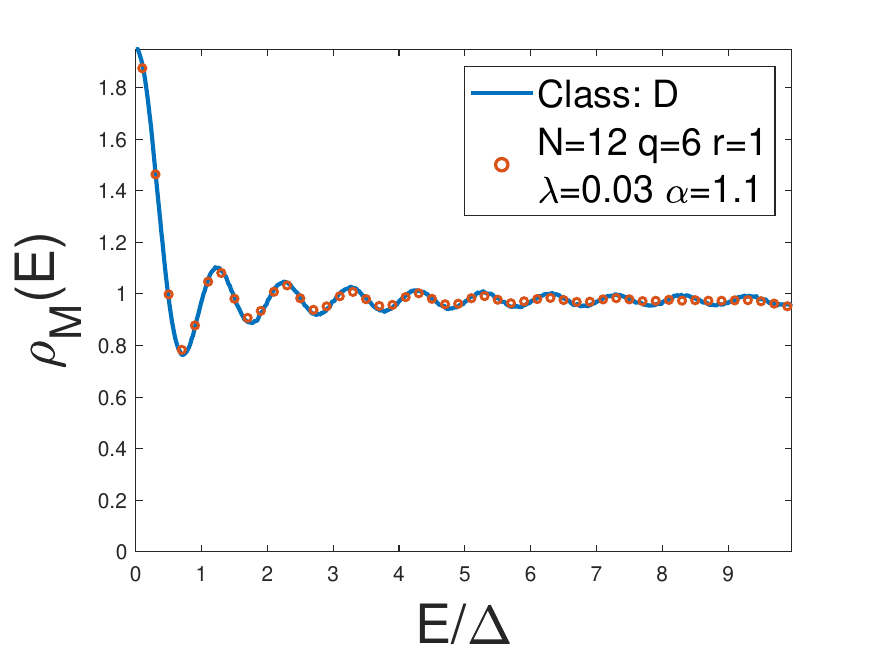}
	\includegraphics[width=4.25cm]{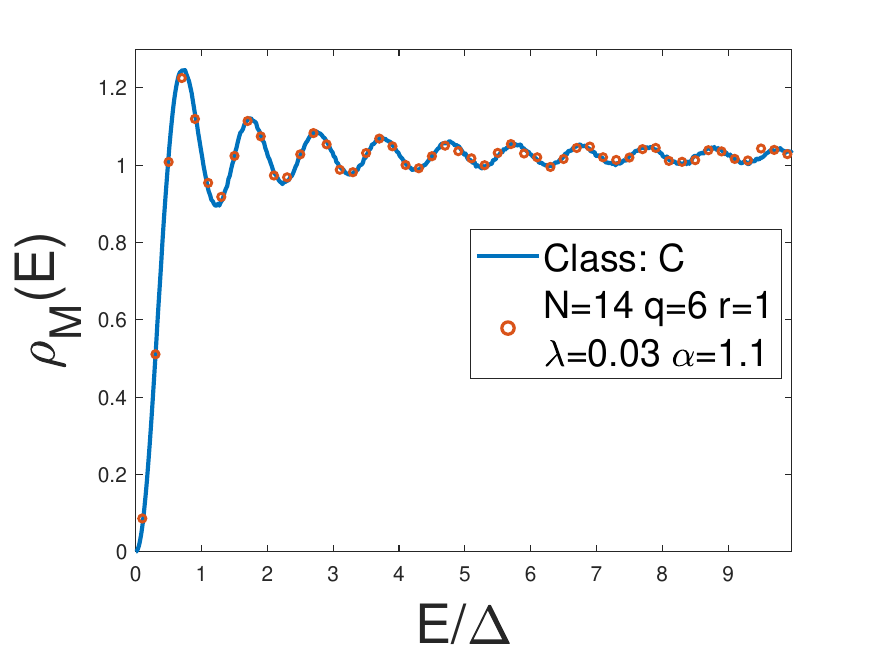}
	\\
	\includegraphics[width=4.25cm]{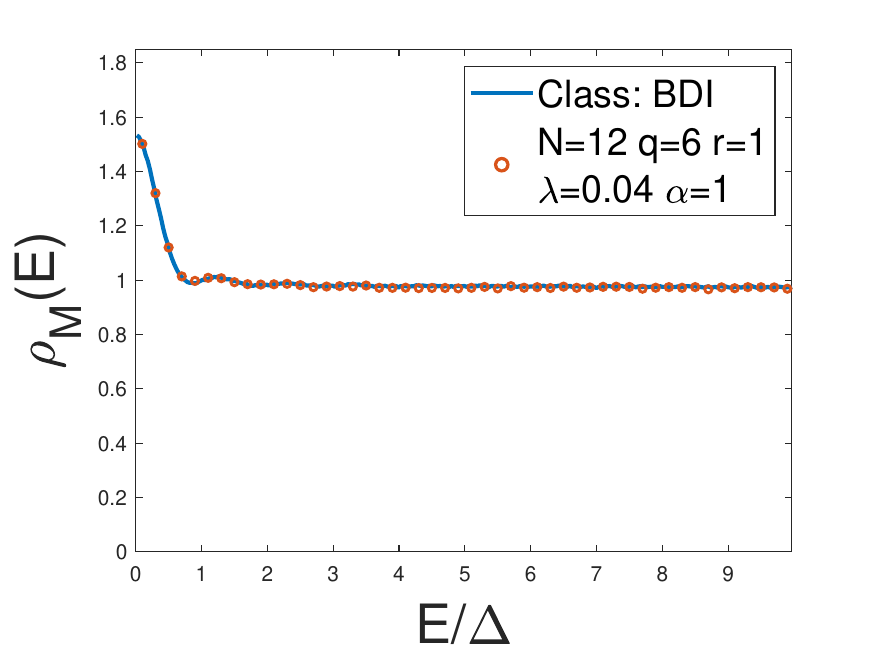}
	\includegraphics[width=4.25cm]{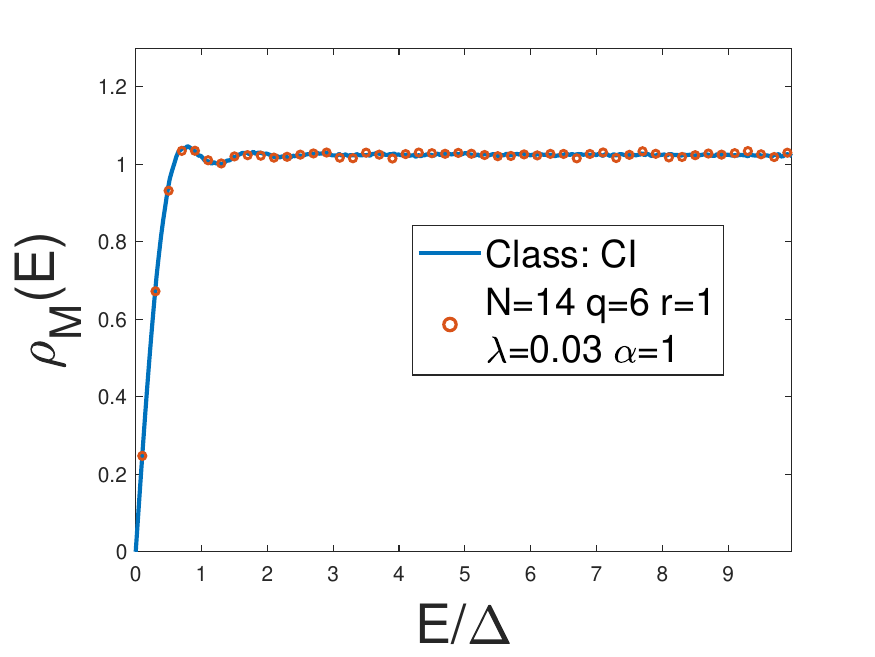}
	\caption{Microscopic spectral density $\rho_M(E)$ in units of the mean level spacing $\Delta$ near $E_0=0$, for the values of the Hamiltonian Eq.~(\ref{eq:hamdef}) indicated in the legend. We find excellent agreement with the RMT result for the predicted universality class.
	}\label{fig:NonDy1}
\end{figure}

The remaining four universality classes (BDI, CI, C, and D) are related to the existence of involutive symmetries that anticommute with the Hamiltonian.
As a result, the spectrum is symmetric around $E_0 = 0$.
Spectral correlations very close to $E_0$, probed by, e.g., the microscopic spectral density~\cite{verbaarschot1993a} expressed in units of the local mean level spacing, or the distribution of eigenvalues closest to $E_0$~\cite{forrester1993NPB,wilke1998PRD,nishigaki1998PRD,damgaard2001PRD,akemann2009PRE,sun2020}, have distinct features that fully characterize the four universality classes.
To illustrate this, in Fig.~\ref{fig:NonDy1}, we compare the microscopic spectral density around $E_0 = 0$ for odd $q/2$ and $r$, and different values of
$N\mod4$ and $\alpha$, corresponding to universality classes BDI ($\alpha=1$, $N\mod4=0$), CI ($\alpha=1$, $N\mod4=2$), D ($\alpha\neq1$, $N\mod4=0$), and C ($\alpha\neq1$, $N\mod4=2$), see Table~\ref{tab:classification_even_odd_odd}.
In all cases, we find excellent agreement with the RMT result. The complementary analysis in terms of the distribution of the eigenvalue closest to $E_0 = 0$, presented in the SM~\cite{SM}, shows a similar agreement.

{\it Traversable wormhole classification.---}%
Having established the symmetry classification of the SYK Hamiltonian Eq.~(\ref{eq:hamdef}), we now study for which parameters ($q$, $r$, $\lambda$, and temperature $T$), this model is related to a traversable wormhole~\cite{gao2016,maldacena2018} in a near AdS$_2$ background~\cite{jackiw1985,teitelboim1983}. First, we note that the traversable wormhole~\cite{maldacena2018} requires a weak intersite coupling $\lambda \ll 1$, and a low temperature $T$, i.e., strong intrasite coupling. The small-$\lambda$ condition is necessary to account rigorously~\cite{gao2016,maldacena2018} for the effect of a double trace deformation coupling the two boundaries in the gravitational path integral.	
In this limit, the holographic relation between the two-site SYK model and the gravity system is established by demonstrating that both models share the same low-energy effective action, which, in this case, is a generalized Schwarzian~\cite{maldacena2018}.
For $q = 4$ and $r = 1$, this program was carried out in Ref.~\cite{maldacena2018}. A distinct feature of the wormhole phase for $r = 1$, confirmed by the numerical solution of the large-$N$ Schwinger-Dyson (SD) equations~\cite{maldacena2016}, is the existence of a gapped ground state at low temperature. Since $\lambda \ll 1$, the gap $E_g \sim \lambda^\gamma$, $\gamma = 2/3 < 1$ for $q=4$, is enhanced with respect to the perturbative result $E_g \sim \lambda$.
Physically, this is interpreted as an enhanced tunneling rate induced by the strong intrasite interactions in the SYK model, and as a traversable wormhole on the gravity side. Generally, a gap $E_g \sim \lambda^\gamma$ with $\gamma <1$ for $\lambda \ll 1$ is a defining feature of the traversable wormhole phase.
Based on this definition, we constrain the previous symmetry classification to the values of $q$ and $r$ for which the SYK model Eq.~(\ref{eq:hamdef}) has an interaction-enhanced gap, i.e., $\gamma < 1$.

\begin{figure}[t!]
	\centering
	\includegraphics[width=7cm]{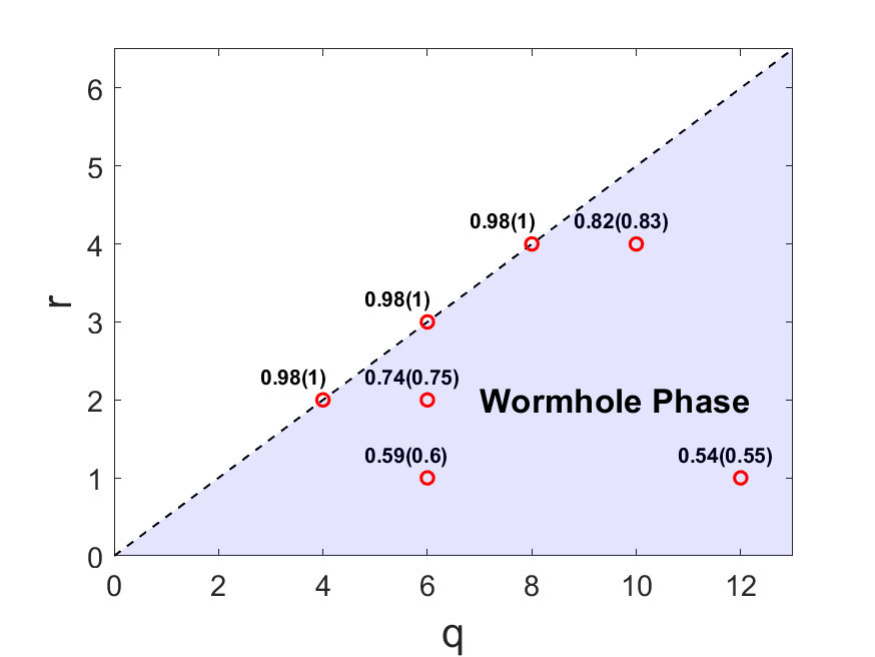}
	\caption{Phase diagram of the Hamiltonian Eq.~(\ref{eq:hamdef}) as a function of $q$ and $r$, obtained from the gap $E_g \propto \lambda^\gamma$. The region of parameters where we expect traversable wormhole physics (purple region) is characterized by $\gamma < 1$ and delimited by the dashed line $q = 2r$. The red circles give $\gamma$ for the different values of $q$ and $r$. The analytic result for $E_g$ [in brackets, see Eq.~(\ref{eq:gap})], is compared to the numerical value obtained from the exponential decay of Green's functions (see text and the SM~\cite{SM}).
	}\label{fig:GapSD}
\end{figure}

In order to proceed, we generalize the results of Ref.~\cite{maldacena2018} by simply replacing $\Delta \equiv 1/q \to r/q$ in the generalized Schwarzian action of Ref.~\cite{maldacena2018}. The resulting gap is given by
\be \label{eq:gap}
E_g  \propto \lambda^{\frac{q}{2(q-r)}}.
\ee	
We have confirmed this scaling with $\lambda$ in the large-$N$ limit by computing $E_g$ numerically from the solution of the SD equations~\cite{maldacena2016}. Results depicted in Fig.~\ref{fig:GapSD} for different $q$ and $r$ show an excellent agreement between the numerical result and analytic prediction $\gamma = {q}/{[2(q-r)]}$. For the technical procedure to solve the SD equations and extract the gap from the Green's function decay, we refer to both Refs.~\cite{maldacena2018,garcia2019} and the SM~\cite{SM}.

As a consequence of Eq.~(\ref{eq:gap}), only SYKs of the form~(\ref{eq:hamdef}) with $q > 2r$ (purple region in Fig.~\ref{fig:GapSD}) can be dual to a traversable wormhole. For $q<2r$ (white region in Fig.~\ref{fig:GapSD}), $\gamma > 1$ and hence there is no tunneling enhancement, so there is no wormhole phase.
The borderline case $q = 2r$ (dashed line in Fig.~\ref{fig:GapSD}) would require further analysis to completely rule out the existence of a wormhole dual.
While we have so far restricted ourselves to the case of identical SYKs, the observation of universality classes D and C requires $\alpha \neq 1$. This is not a problem because wormhole features are not qualitatively affected provided that $\alpha$ is sufficiently close to one~\cite{maldacena2018,garcia2019}. Most importantly, the condition $q > 2r$ does not restrict the possible symmetry classes and all six occur for either $\alpha =1$ or $\alpha \approx 1$. Finally, we note that another feature associated with a traversable wormhole, namely, the existence of a first-order phase transition in the free energy, also occurs in our SYK setting, see the SM~\cite{SM}.

In conclusion, based on the relation between a two-site SYK model at low temperature and weak intersite coupling, we have identified AdS$_2$ traversable wormholes belonging to six universality classes: A, AI, BDI, CI, C, and D.
Wormholes with symplectic symmetry (classes AII, CII, and DIII) are conspicuously missing. Moreover, enhanced tunneling that is a signature of wormhole physics only occurs for $q > 2r$, see Eq.~(\ref{eq:gap}).
A natural extension of this work is the symmetry classification of coupled non-Hermitian SYKs, whose gravity dual are Euclidean and Keldysh wormholes~\cite{maldacena2018,garcia2021,garcia2022e,kawabata2022,garcia2023PT}.

\let\oldaddcontentsline\addcontentsline% Store \addcontentsline
\renewcommand{\addcontentsline}[3]{}% Make \addcontentsline a no-op

\acknowledgments{
\textit{Acknowledgments.---}%
A.\ M.\ G.\ G.\ and Y.\ C.\ acknowledge support from the National Natural Science Foundation of China (NSFC): Individual Grant number 12374138, Research Fund for International Senior Scientists number 12350710180, and National Key R\&D Program of China (Project ID: 2019YFA0308603). A.\ M.\ G.\ G.\ acknowledges support from a Shanghai talent program. L.\ S.\ was supported by a Research Fellowship from the Royal Commission for the Exhibition of 1851 and by Fundação para a Ciência e a Tecnologia (FCT-Portugal) through Grant No.\ SFRH/BD/147477/2019. J.\ J.\ M.\ V.\ acknowledges support from U.S.\ DOE Grant No.\ DE-FAG-88FR40388.
}

\bibliography{librarynh.bib}

\let\addcontentsline\oldaddcontentsline% Restore \addcontentsline

\onecolumngrid

%SUPPLEMENTAL MATERIAL

\clearpage

\setcounter{table}{0}
\renewcommand{\thetable}{S\arabic{table}}%
\setcounter{figure}{0}
\renewcommand{\thefigure}{SM\arabic{figure}}%
\setcounter{equation}{0}
\renewcommand{\theequation}{S\arabic{equation}}%
\setcounter{page}{1}
\renewcommand{\thepage}{SM-\arabic{page}}%
\setcounter{secnumdepth}{3}
\setcounter{section}{0}
\renewcommand{\thesection}{\arabic{section}}%
\setcounter{subsection}{0}
\renewcommand{\thesubsection}{\arabic{section}.\arabic{subsection}}%

\begin{center}\Large{
		\textit{Supplemental Material for}\\
		\textbf{Sixfold way of traversable wormholes in the Sachdev-Ye-Kitaev model}\\
		by Antonio M.\ Garc\'\i a-Garc\'\i a, Lucas S\'a, Jacobus J.\ M.\ Verbaarschot and Can Yin
		%\begin{CJK*}{UTF8}{gbsn}(殷灿)\end{CJK*}
	}
\end{center}

{
	\hypersetup{linkcolor=black}
	\tableofcontents
}

\section{Details on the symmetry classification of the two-site SYK model}

In this appendix, we give a detailed step-by-step classification of two-site coupled SYK
Hamiltonians. While in the main text we restricted our attention to even $N$ only, here we present a simultaneous unified treatment of both even and odd $N$. Note that considering an odd number of Majorana fermions in each site of the model is not problematic, since the total number of Majoranas, $2N$, is always even. In any case, we find below that for odd $N$ only classes A and AI arise (no new classes) and it is, therefore, of limited interest. Moreover, we have restricted our classification to even $q$ only.
For a single SYK, an odd $q$ operator is the supercharge $Q$ with supersymmetric Hamiltonian $H = Q^2$~\cite{li2017,fu2018,kanazawa2017,garcia2018a}, but it
is not clear how to extend this to two coupled SYK's.

As discussed in the main text, the symmetry classification of the two-site Hamiltonian,
\begin{equation}
\begin{split}\label{eq:hami}
&H=H_0+\lambda H_I,\\
&H_0=H_L+\alpha (-1)^{q/2} H_R,\\
&H_L=-\i^{q/2} \sum_{i_1<\cdots<i_q}^{N} J_{i_1\cdots i_q} \, \psi^L_{i_1} \cdots \, \psi^L_{i_q},
\\
&H_R=-\i^{q/2}\sum_{i_1<\cdots<i_q}^{N} J_{i_1\cdots i_q} \, \psi^R_{i_1} \cdots \, \psi^R_{i_q},
\\
&H_I=\i^r \frac{N^{1-r}}{r} \left(\sum_{i=1}^{N}\psi^L_{i}\psi^R_{i}\right)^r,
\end{split}
\end{equation}
is based upon the action of antiunitary symmetries built from the complex-conjugation operator $K$ and the following unitary operators: the left and right parity,
\begin{equation}
S_L=\i^{N(N-1)/2}\prod_{i=1}^N\sqrt{2}\psi_i^L,
\qquad
S_R=\i^{N(N+1)/2}\prod_{i=1}^N\i\sqrt{2}\psi_i^R,
\end{equation}
the total parity,
\begin{equation}
S=(-1)^{N^2/2}S_LS_R,
\end{equation}
and the (exponential of the) spin,
\begin{equation}
Q=(-1)^{N/2(N/2-1)}\exp\left\{-\frac{\pi}{4}\sum_{i=1}^N\psi_i^L\psi_i^R\right\}
=(-1)^{N/2(N/2-1)}\prod_{i=1}^N\frac{1}{\sqrt{2}}\(1-2\psi_i^L\psi_i^R\).
\end{equation}
These unitary operators square to
\begin{align}
\label{eq:square_unitaries}
S_L^2=S_R^2=S^2=+1,\qquad
Q^2=S,
\end{align}
and satisfy the commutation relations
\begin{align}
\label{eq:comm_SLSR}
&S_L S_R =(-1)^{N} S_R S_L,
\\
\label{eq:comm_QSL}
&Q S_L = (-1)^{N^2/2}S_L Q^{-1}=(-1)^N S_R Q= (-1)^{-N^2/2}S S_L Q,
\\
\label{eq:comm_QSR}
&Q S_R = (-1)^{N^2/2}S_R Q^{-1}=S_L Q= (-1)^{-N^2/2}S S_R Q,
\\
\label{eq:comm_QS}
&QS=SQ.
\end{align}
They act on Majorana fermions as
\begin{align}
\label{eq:action_SL_psi}
&S_L \psi_i^L S_L^{-1} = -(-1)^N\psi_i^L,
&&S_L \psi_i^R S_L^{-1} = (-1)^N\psi_i^R,
\\
\label{eq:action_SR_psi}
&S_R \psi_i^L S_R^{-1} = (-1)^N\psi_i^L,
&&S_R \psi_i^R S_R^{-1} = -(-1)^N\psi_i^R,
\\
\label{eq:action_S_psi}
&S \psi_i^L S^{-1} = -\psi_i^L,
&&S \psi_i^R S^{-1} = -\psi_i^R,
\\
\label{eq:action_Q_psi}
&Q \psi_i^L Q^{-1} = \psi_i^R,
&&Q \psi_i^R Q^{-1} = -\psi_i^L.
\end{align}

In order to fix the action of the antiunitary complex-conjugation operator $K$, which is basis dependent, we choose the left Majorana fermions to be represented by real symmetric matrices and the right Majorana fermions to be purely imaginary and antisymmetric,
\begin{align}
\label{eq:action_K_psi}
K \psi_i^L K^{-1} = \psi_i^L,
\qquad
K \psi_i^R K^{-1} = -\psi_i^R.
\end{align}
This action can be implemented in practice by representing the Majoranas with Pauli strings,
\begin{align}
\psi_i^L = \frac{1}{\sqrt{2}} (\sigma_3)^{\otimes(i-1)}\otimes \sigma_1,\\
\psi_i^R = \frac{1}{\sqrt{2}} (\sigma_3)^{\otimes(i-1)}\otimes \sigma_2,
\end{align}
where $\sigma_{1,2,3}$ are the standard Pauli matrices.  Furthermore, in this basis, the unitary operators $S_{L,R}$ are antidiagonal, $Q$ and $S$ are diagonal satisfying,
\begin{align}
\label{eq:comm_K_unitaries}
K S_L= (-1)^{N(N-1)/2}S_L K,
\qquad
K S_R =(-1)^{N(N+1)/2}S_R K,
\qquad
K S = S K,
\qquad
K Q=Q^{-1} K=S  QK.
\end{align}
For later use, we also need the commutation relation of the complex-conjugation operator with the projectors $\mathbb{P}$ into sectors of fixed eigenvalues of $S_{L,R}$, $S$, and $Q$, defined respectively as,
\begin{align}
\label{eq:PLR}
\mathbb{P}_{L,R}^{s_{L,R}}&=\frac{1}{2}\(\id+s_{L,R}S_{L,R}\),
\\
\label{eq:PS}
\mathbb{P}_{S}^{s}&=\frac{1}{2}\(\id+sS\),
\\
\label{eq:PQ}
\mathbb{P}_{Q}^{k}&=\frac{1}{4}\(\id+\frac{Q}{k}+\frac{Q^2}{k^2}+\frac{Q^3}{k^3}\),
\end{align}
where $s_{L,R}=\pm1$, $s=\pm1$ and $k=\pm1,\pm\i$ denote the eigenvalues of $S_{L,R}$, $S$, and $Q$, respectively.
Note that $\mathbb{P}_{Q}^{k}$ can also be written as
\begin{equation}
\label{eq:PQ-alt}
\mathbb{P}_{Q}^{k}=\frac{1}{4}\(\id+s S\)\(\id+\frac{Q}{k} \).
\end{equation}
In our basis, the projectors $\mathbb{P}_S$ and $\mathbb{P}_Q$ are always real, since they are diagonal with entries either zero or one, and, hence, they commute with $K$. The projector $\mathbb{P}_{L,R}$ commutes with $K$ only when $S_{L,R}$ is real, i.e., when $N\mod4=0,1$ for $S_L$ and $N\mod4=0,3$ for $S_R$.
A basis independent proof of the reality of $\mathbb{P}_{Q}^{k}$ follows from
$ K Q = Q^{-1} K$, $Q^4=1$, and $k^4=1$ resulting in,
\be
K \mathbb{P}_{Q}^{k}=\frac 14 \(\id +\frac{Q^{-1}}{k^*}+\frac{Q^{-2}}{{k^*}^2}
+\frac{Q^{-3}}{{k^*}^3}+\frac{Q^{-4}}{{k^*}^4} \)K=\mathbb{P}_{Q}^{k} K.
\ee
Using Eqs.~(\ref{eq:action_SL_psi})--(\ref{eq:action_K_psi}), we find the transformation relations of the Hamiltonian under the unitary and antiunitary transformations, as stated in the main text:
\begin{align}
\label{eq:action_SL_H}
&S_L \(H_L+\alpha (-1)^{q/2}H_R\) S_L^{-1} = H_L+\alpha (-1)^{q/2} H_R,
&&S_L H_I S_L^{-1} = (-1)^r H_I,
\\
\label{eq:action_Q_H}
&Q \(H_L+\alpha (-1)^{q/2}H_R\) Q^{-1} = \alpha(-1)^{q/2} \(H_L+\alpha^{-1} (-1)^{q/2}H_R\),
&&Q H_I Q^{-1} = H_I,
\\
\label{eq:action-K-H}
&K \(H_L+\alpha (-1)^{q/2}H_R\) K^{-1} = (-1)^{q/2} \(H_L+\alpha (-1)^{q/2}H_R\),
&&K H_I K^{-1} = H_I.
\end{align}
The action of $S_R$ on $H$ is the same as that of $S_L$. Note that, consequently, $H$ commutes with $S$ for all $N$, $q$, and $r$, i.e., the total fermionic parity is always conserved as it should. Moreover, $Q$ only implements a symmetry or involution of the Hamiltonian if $\alpha=1$, that is, if there is left-right symmetry of the Hamiltonian.

The symmetry classification follows from a careful and systematic evaluation of Eqs.~(\ref{eq:square_unitaries})--(\ref{eq:comm_QS}), (\ref{eq:comm_K_unitaries})--(\ref{eq:PQ}), and (\ref{eq:action_SL_H})--(\ref{eq:action-K-H}). Particular care must be taken with the block structure of the Hamiltonian defined by its commuting unitary symmetries~\cite{Kieburg:2014eca}, since time-reversal, particle-hole, and chiral symmetries acting within each such block should be well defined, i.e., they must commute with the projector into a given block. Furthermore, if there are multiple commuting unitaries, we must also check whether they commute with each other and define a common eigenbasis.
The results for even $N$ are summarized in Tables~\ref{tab:classification_even_even_even}--\ref{tab:classification_even_odd_odd} in the main text, while those for odd $N$ are summarized in Tables~\ref{tab:classification_odd_even} and \ref{tab:classification_odd_odd}.

We now look at all these cases in detail, starting with the left-right symmetric case ($\alpha=1$).

\subsection{Left-right symmetric case: $\alpha = 1$}

\paragraph*{\textbf{Even $\bm{q/2}$, even $\bm{r}$.}}
The operators $Q$, $S_L$, and $S_R$ all commute with $H$, but do not necessarily commute with each other. The operators $S_L$ and $S_R$ commute if $N$ is
even and anticommute if $N$ is odd,
hence there are blocks of simultaneously fixed eigenvalues of $S_L$ and $S_R$ in the former, and only blocks of fixed eigenvalue of $S$ in the latter.
The spin operator $Q$ may further split some of the blocks. Time-reversal symmetry is implemented by the antiunitary operator $T=K$, which always squares to one.
\begin{itemize}
\item \textit{Even $N$.} The parity operators $S_L$ and $S_R$ commute and there are four blocks with fixed left and right parity eigenvalues. The spin operator $Q$ has eigenvalues $k=\pm1,\pm\i$ and $Q^2=S$. If we are in a block with $s=+1$ (i.e., $k=\pm1$), it follows from Eqs.~(\ref{eq:comm_QSL}) and (\ref{eq:comm_QSR}) that $Q$ commutes with $S_{L,R}$ and, therefore, the eigenvalue of $Q$ is a good quantum number for the $s=1$ blocks, splitting them into two sub-blocks each, with eigenvalues of $Q$ equal to $k=\pm1$.
  In the two blocks with $s=-1$, the spin operator $Q$ does not commute
  with $S_{L,R}$, it is not a good quantum number, i.e., states with $k=\pm\i$ are linear superpositions of states with $s_{L,R}=\pm1$ and vice-versa. In total, we thus have six blocks of $H$.
Because $\mathbb{P}_Q$ is diagonal and real,
  the time reversal operator $T$ always commutes with this operator. If $N\mod4=0$, $S_{L,R}$ is real and $T$ also commutes with the projector $\mathbb{P}_{L,R}$, leading to all six blocks having GOE statistics. Since there is no other antiunitary symmetry, they belong to class AI, see Table~\ref{tab:classification_even_even_even}. If $N\mod4=2$, $s_{L,R}$ is imaginary, and $T$ maps a state in the block with $s_{L,R}=+1$ into a state in the block with $s_{L,R}=-1$. The blocks with $s=+1$ thus belong to class A and have GUE statistics. The blocks with $s=-1$ can be either split by $S_{L}=-S_R$ or by $Q$ (but, importantly, not both simultaneously). Since this choice cannot affect the level statistics and the projector $\mathbb{P}_Q$ commutes with $T$, these two blocks belong to class AI and display GOE statistics, see Table~\ref{tab:classification_even_even_even}.

\item \textit{Odd $N$.} None of $S_L$, $S_R$, or $Q$ commute with each other and, therefore, do not define a common eigenbasis. We choose the blocks to have a fixed eigenvalue of $Q$, which leads to the partition of $H$ into four blocks of eigenvalues $k=\pm1,\pm\i$. Since the projector $\mathbb{P}_Q$ is always real, all four blocks have GOE level statistics. Since, again, there is no other antiunitary symmetry, all blocks belong to class AI, see Table~\ref{tab:classification_odd_even}.
\end{itemize}

Before proceeding to the next case, we remark that, from Eqs.~(\ref{eq:action_SL_H}) and (\ref{eq:action-K-H}), it follows that the antiunitary operator $T'=S_L K$ also acts as a TRS of $H$ and squares to $T'^2=(-1)^{N(N-1)/2}$. At first sight, it seems we could have GSE level statistics (class AII), in contradiction with our findings above. However, we show in the following that whenever $T'^2=-1$, it does not act within a single block of $H$ and, therefore, does not implement a well-defined TRS.
\begin{itemize}
	\item \textit{Even $N$.} The blocks of $H$ are labeled by the eigenvalues of $S_L$, $S_R$, and if $S_L=S_R$ by $Q$. The commutation relation of $T'$ with the projector into blocks of $S_L$ is
	\begin{equation}
	T'\mathbb{P}_L^{s_L}
	=\frac{1}{2}S_LK\(\id +s_L S_L\)
	=\frac{1}{2}\(\id +(-1)^{N/2}s_LS_L\)S_LK
	=\mathbb{P}_L^{(-1)^{N/2}s_L}T',
	\end{equation}
	while, for the projector into blocks of fixed $Q$, it is
	\begin{equation}
	T'\mathbb{P}_Q^{k}
	=\frac{1}{4}S_LK\(\id+\frac{Q}{k}+\frac{Q^2}{k^2}+\frac{Q^3}{k^3}\)
	=\frac{1}{4}\(\id+S \frac{Q}{k}+\frac{Q^2}{k^2}+S \frac{Q^3}{k^3}\)S_LK
	=\mathbb{P}_Q^{sk}T'.
	\end{equation}
	It follows that $T'$ only acts within single blocks of $H$ when $N\mod4=0$ and $s=+1$. In that case, $T'^2=+1$ and the blocks belong to class AI, in agreement with Table~\ref{tab:classification_even_even_even}. Indeed, in these cases, we have that $T'=S_L Q T$, i.e., the product of the TRS operator $T$ and the conserved quantities (commuting unitary symmetries) $S_L$ and $Q$; the classification must, therefore, agree with the one elaborated on above. We also see that when $N\mod4=2$ and, consequently, $T'^2=-1$, $T'$ always connects blocks of opposite $s_L$.

      \item \textit{Odd $N$.} The blocks of $H$ are labeled by $S$ and $Q$. The commutation relation of $T'$ and $S$ is
	\begin{equation}
	T'\mathbb{P}_S^{s}
	=\frac{1}{2}S_LK\(\id +s S\)
	=\frac{1}{2}\(\id-sS\)S_LK
	=\mathbb{P}_L^{-s}T',
	\end{equation}
	$T'$ always connects blocks with opposite total parity and is not
        a symmetry of the Hamiltonian.
\end{itemize}

\paragraph*{\textbf{Even $\bm{q/2}$, odd $\bm{r}$.}}
The Hamiltonian $H$ again commutes with $Q$ and the total parity $S$, but not with $S_L$ and $S_R$ individually. Because $Q$ and $S$ always commute, they define a common eigenbasis. Since the blocks with fixed eigenvalue of $S$ split into two independent sub-blocks of fixed eigenvalue of $Q$ each, it is enough to consider the latter. The four blocks are labeled by the eigenvalues of $Q$, $k=\pm1,\;\pm\i$.
TRS is again implemented by $T=K$ and, because the projector $\mathbb{P}_Q$ is always real, all four blocks of $H$ display GOE statistics. Since there are no other antiunitary symmetries, the four blocks belong to class AI regardless of $N$, see Table~\ref{tab:classification_even_even_odd} and \ref{tab:classification_odd_even}.

\paragraph*{\textbf{Odd $\bm{q/2}$, even $\bm{r}$.}}
For odd $q/2$, $Q$ is no longer a symmetry of $H$ (because of the $(-1)^{q/2}$ factor in front of $H_R$), but $S_L$ and $S_R$ still are. The TRS operator is now given by
the antiunitary operator $T=QK$, which always squares to $+1$ because $Q^*=Q^{-1}$. This is the only antiunitary symmetry.
\begin{itemize}
	\item \textit{Even $N$:} $S_L$ and $S_R$ commute and $H$ splits into four blocks. Combining Eqs.~(\ref{eq:comm_QSL}), (\ref{eq:comm_K_unitaries}), and (\ref{eq:PLR}), we have that
	\begin{align}
	T\mathbb{P}_L^{s_L}
	=\frac{1}{2}QK\(\id +s_L S_L\)
	=\frac{1}{2}\(\id +(-1)^{N/2}s_L S  S_L\)QK
	=\frac{1}{2}\(\id +(-1)^{N/2}s_R  S_L\)QK
	=\mathbb{P}_L^{(-1)^{N/2}s s_L}T,
	\end{align}
	and similarly for $\mathbb{P}_R$,
\be
	T\mathbb{P}_R^{s_R}
	=\mathbb{P}_R^{(-1)^{N/2}s s_R}T.
\ee
         We conclude
        that for $s (-1)^{N/2} =1$,
        $T$ commutes with both projectors and acts within each block of $H$, so that these blocks belong to class AI. On the contrary, if
        $s (-1)^{N/2} =-1$, the time-reversal operator connects a $s_L=+1$ block with a $s_L=-1$ block (and similarly for $s_R$) and these blocks are in class A. See Table~\ref{tab:classification_even_odd_even} for a summary of all cases.

      \item \textit{Odd $N$.}The parity operators $S_L$ and $S_R$ anticommute and do not define a common eigenbasis. We can split the Hamiltonian into two blocks of conserved $S$. Because $S$ commutes with $Q$ and is diagonal with
        real eigenvalues $\pm 1$ in the convention (\ref{eq:action_K_psi}), the TRS operator $T$ commutes with the projector $\mathbb{P}_S$ and, hence, both blocks belong to class AI, see Table~\ref{tab:classification_odd_odd}.
\end{itemize}

\paragraph*{\textbf{Odd $\bm{q/2}$, odd $\bm{r}$.}}
In this case, $S$ is the only commuting unitary symmetry of $H$, which splits into two blocks with $S=\pm1$. The operator $T=QK$ is again an antiunitary symmetry that commutes with $H$ (TRS), with $T^2=+1$. However, there is now a second antiunitary operator that {\it anticommutes} with $H$ (PHS), $C=S_L K$, which satisfies $C^2=S_LKS_LK=(-1)^{N(N-1)/2}$. Combining Eqs.~(\ref{eq:comm_SLSR}), (\ref{eq:comm_K_unitaries}), and (\ref{eq:PS}), the commutation relation of $C$ and the projector $\mathbb{P}_S$ is:
\begin{align}
\label{eq:comm_C_PS}
C\mathbb{P}_S^s
=\frac{1}{2}S_LK\(\id+s S\)
=\frac{1}{2}\(\id +(-1)^N S\)S_LK
=\mathbb{P}_S^{s(-1)^N}C.
\end{align}
\begin{itemize}
	\item \textit{Even $N$.} The PHS operator $C$ acts within each block and there is PHS, in addition to TRS. If $N\mod4=0$, $C^2=+1$ and both blocks belong to class BDI, while if $N\mod4=2$, $C^2=-1$ and both blocks belong to class CI, see Table~\ref{tab:classification_even_odd_odd}.
	\item \textit{Odd $N$.} The PHS operator $C$
anticommutes with $S$ and
          connects different parity blocks. Consequently, there is no particle-hole symmetry. The two blocks belong to class AI, see Table~\ref{tab:classification_odd_odd}.
\end{itemize}

\subsection{Left-right asymmetric case: $\alpha\neq 1$}
We now turn to the left-right asymmetric case, $\alpha\neq 1$, for which $Q$ does not act as a symmetry of the Hamiltonian. As a consequence, for even $q/2$, $Q$ is not a unitary symmetry and the parity blocks are not split by it, while for odd $q/2$, there is no antiunitary TRS.

\begin{table}[t]
	\caption{Symmetry classification of the two-site SYK Hamiltonian for odd $N$ and even $q/2$. The results are the same for $N\mod4=1$ and $3$ and are independent of $r$. Each line corresponds to a block of the Hamiltonian, labeled by the eigenvalues of the conserved quantities $S$ and $Q$. For each of the four blocks, we have given its dimension and the symmetry class to which it belongs for the left-right symmetric ($\alpha=1$) and asymmetric ($\alpha\neq 1$) cases.}
	\label{tab:classification_odd_even}
	\begin{tabular}{@{}Sc Sc Sc Sc Sc@{}}
		\toprule
		$S$                   & $Q$   & Dimension             & $\alpha=1$ & $\alpha\neq1$         \\ \midrule
		\multirow{2}{*}{$+1$} & $+1$  & $(2^N+2^{(N+1)/2})/4$ & AI         & \multirow{2}{*}{AI}   \\
		                      & $-1$  & $(2^N-2^{(N+1)/2})/4$ & AI         &                       \\ \midrule
		\multirow{2}{*}{$-1$} & $+\i$ & $(2^N-2^{(N+1)/2})/4$ & AI         & \multirow{2}{*}{AI}   \\
	                          & $-\i$ & $(2^N+2^{(N+1)/2})/4$ & AI         &                       \\ \bottomrule
	\end{tabular}
	
	\caption{Same as Table~\ref{tab:classification_odd_even}, but for odd $q/2$. There are two blocks labeled by the eigenvalues of $S$.}
	\label{tab:classification_odd_odd}
	\begin{tabular}{@{}Sc Sc Sc Sc@{}}
		\toprule
		$S$      & Dimension & $\alpha=1$ & $\alpha\neq1$ \\ \midrule
		$+1$     & $2^N/2$ & AI         & A             \\
		$-1$     & $2^N/2$ & AI         & A             \\
		\bottomrule
	\end{tabular}
\end{table}

\paragraph*{\textbf{Even $\bm{q/2}$, even $\bm{r}$.}} The Hamiltonian $H$ conserves both $S_L$ and $S_R$, which commute for even $N$ and anticommute for odd $N$. The resulting blocks are not split by $Q$ as before.
\begin{itemize}
	\item \textit{Even $N$}. There are four blocks of conserved $S_L$ and $S_R$. If $N\mod4=0$, $T=K$ commutes with the projectors $\mathbb{P}_{L,R}$ and the four blocks belong to class AI (GOE statistics), if $N\mod4=2$, $T$ does not commute with $\mathbb{P}_{L,R}$ and the four blocks belong to class A (GUE statistics), see Table~\ref{tab:classification_even_even_even}.
	\item \textit{Odd $N$.} There are two blocks of conserved $S$. Since $T$ always commutes with $S$, both belong to class AI (GOE statistics), see Table~\ref{tab:classification_odd_even}.
\end{itemize}

\paragraph*{\textbf{Even $\bm{q/2}$, odd $\bm{r}$.}} There are two blocks of conserved $S$. They are not further split by $Q$. Since $T=K$ always commutes with $S$, both belong to class AI (GOE statistics), irrespective of $N$, see Tables~\ref{tab:classification_even_even_odd} and \ref{tab:classification_odd_even}.

\paragraph*{\textbf{Odd $\bm{q/2}$, even $\bm{r}$.}} There are four blocks of conserved $S_{L}$ and $S_R$. Since there is no antiunitary symmetry, all four belong to class A (GUE statistics), irrespective of $N$, see Tables~\ref{tab:classification_even_odd_even} and \ref{tab:classification_odd_odd}.

\paragraph*{\textbf{Odd $\bm{q/2}$, odd $\bm{r}$.}} There are two blocks of conserved $S$, no TRS $T$, but the same PHS $C=S_L K$ as in the case of $\alpha=1$.

\begin{itemize}
	\item \textit{Even $N$.} The PHS $C$ commutes with the projector $\mathbb{P}_{S}$. If $N\mod4=0$, $C^2=+1$ and the two blocks belong to class D, if $N\mod4=2$, $C^2=-1$ and the two blocks belong to class C, see Table~\ref{tab:classification_even_odd_odd}.
	\item \textit{Odd $N$.} The PHS $C$ does not commute with the projector $\mathbb{P}_{S}$ and the two blocks belong to class A, see Table~\ref{tab:classification_odd_odd}.
\end{itemize}

\subsection{Asymmetric coupling Hamiltonian: GSE level statistics}

The absence of GSE statistics (classes AII, CII, and DIII) is intimately connected with the left-right symmetry of the coupling Hamiltonian $H_I$. In the following, we will show that, if we consider an asymmetric coupling Hamiltonian $H_I'$ with a different number of left and right Majoranas, it is possible to obtain classes with GSE level statistics. However, such a coupling term is nonstandard and we have no physical reason to propose it. Since it is also unclear whether this model yields wormhole solutions, we consider both the sixfold way presented in the main text and the previous sections as more fundamental.

The left-right asymmetric coupling Hamiltonian is
\begin{equation}
H_I'=\i^{(r+s)/2} \frac{N^{1-(r+s)/2}}{(r+s)/2}\sum_{\substack{i_1<\cdots<i_r\\j_1<\cdots<j_s}}^N\psi_{i_1}^L \cdots \psi_{i_r}^L \psi_{j_1}^R\cdots \psi_{j_s}^R,
\end{equation}
with $r$ and $s$ two unequal integers of the same parity. The Hamiltonian
$H_0$ and the remaining parameters are the same as before. The total Hamiltonian is $H'=H_0+\lambda H_I'$.

We consider again the action of $Q$, $S_{L,R}$, $S$ and $K$ on the Hamiltonian. The operator
$Q$ is not a symmetry of $H'$ because it transforms an interaction with $r$ left fermions and $s$ right fermions into one with $s$ left fermions and $r$ right fermions. Because the Hamiltonian is still bosonic ($q$ and $r+s$ are even), $S$ again commutes with $H'$. It suffices to consider the action of $S_{L,R}$ and $K$:
\begin{align}
\label{eq:GSE_action_SL_H}
&S_L \(H_L+\alpha (-1)^{q/2}H_R\) S_L^{-1} = H_L+\alpha (-1)^{q/2} H_R,
&&S_L H_I' S_L^{-1} = (-1)^r H_I',
\\
\label{eq:GSE_action_SR_H}
&S_R \(H_L+\alpha (-1)^{q/2}H_R\) S_R^{-1} = H_L+\alpha (-1)^{q/2} H_R,
&&S_R H_I' S_R^{-1} = (-1)^s H_I',
\\
\label{eq:GSE_action_K_H}
&K \(H_L+\alpha (-1)^{q/2}H_R\) K^{-1} = (-1)^{q/2} \(H_L+\alpha (-1)^{q/2}H_R\),
&&K H_I' K^{-1} = (-1)^{(r-s)/2}H_I'.
\end{align}

The symmetry classification of $H'$ follows from a systematic analysis of Eqs.~(\ref{eq:square_unitaries})--(\ref{eq:comm_QS}), (\ref{eq:comm_K_unitaries})--(\ref{eq:PQ}), and (\ref{eq:GSE_action_SL_H})--(\ref{eq:GSE_action_K_H}), and depends on $N\mod4$, the parity of $q/2$, $r$, and $(r+s)/2$, and whether $\alpha=1$ or $\alpha\neq1$. Since this is not the focus of our paper, we will not carry it out in full generality and address only the cases for which GSE statistics can arise: even $q/2$, odd $r$ and even $(r+s)/2$ (e.g., $r=3$, $s=1$), $\alpha=1$, and even $N$. The parity operator $S_L$ does not commute with $H_I'$, and $H'$ splits into two blocks of fixed $S=\pm1$. The complex conjugation operator
$K$ commutes with $H_0$ but anticommutes with $H_I'$, thus not defining a symmetry. However, we have a TRS operator $T=S_L K$, which, as we have seen above, squares to $T^2=(-1)^{N(N-1)/2}$. There is no further antiunitary symmetry (PHS, $C$). Moreover, using Eq.~(\ref{eq:comm_C_PS}), we find that $T$ always commutes with the projector into blocks of fixed $S$ and acts as a TRS inside a single block of $H'$. If $N\mod4=0$, $T^2=+1$, the two blocks belong to class AI, and its eigenvalues display GOE level statistics. If $N\mod4=2$, $T^2=-1$, the two blocks belong to class AII, and its eigenvalues are doubly degenerate (Kramer's degeneracy) and display GSE level statistics.

\section{Confirmation of the symmetry classification by a level statistics analysis}

In this appendix, we study level correlations as a function of the parameters of the Hamiltonian. In Sec.~\ref{sec:2.1}, we study the bulk level correlations using the spacing ratio distribution, and in Sec.~\ref{sec:2.2}, we study the distribution
of the smallest nonzero eigenvalue for the cases with a reflection symmetric spectrum. In all cases we find agreement with the predicted RMT behavior.

\subsection{Bulk level correlations: spacing ratio distribution}
\label{sec:2.1}

\begin{figure}[t]
	\centering
	\includegraphics[width=\textwidth]{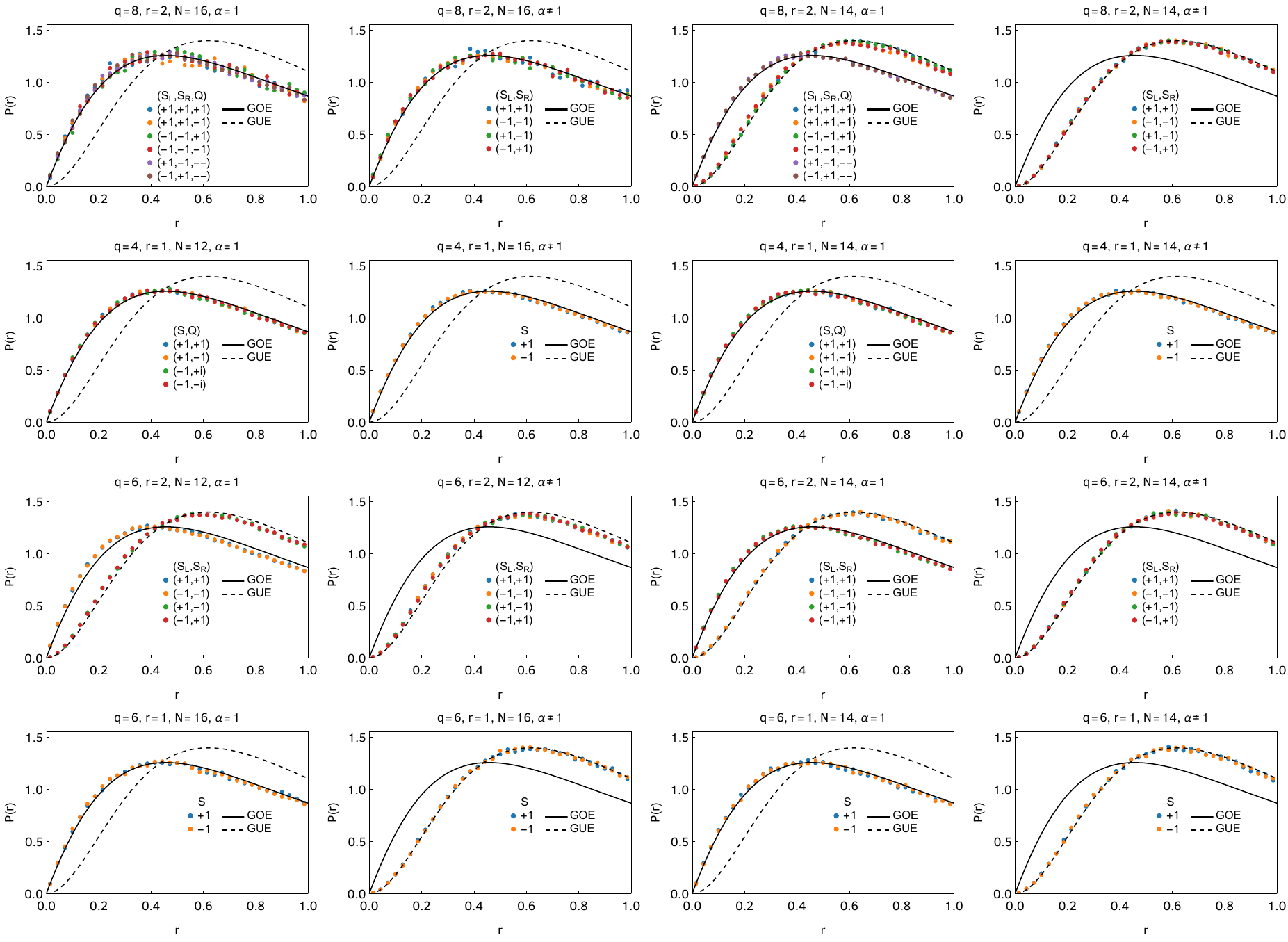}
	\caption{Spacing ratio distribution $P(r)$ of the blocks of $H$ for even $N$ and the different values of the parameters as indicated in each plot above. For all cases $\lambda=0.15$ and we take $\alpha=1.1$ whenever $\alpha\neq1$ is indicated. The colored dots represent the numerical results obtained from exact diagonalization for the different blocks of the Hamiltonian, while the black (full and dashed)
          curves give the surmise for the GOE and GUE, Eq.~(\ref{eq:surmise}). We find excellent agreement with the RMT predictions for all cases, see Tables~\ref{tab:classification_even_even_even}--\ref{tab:classification_even_odd_odd}.
	}
	\label{fig:SM_LevelStats_EvenN}
\end{figure}

\begin{figure}[t]
	\centering
	\includegraphics[width=\textwidth]{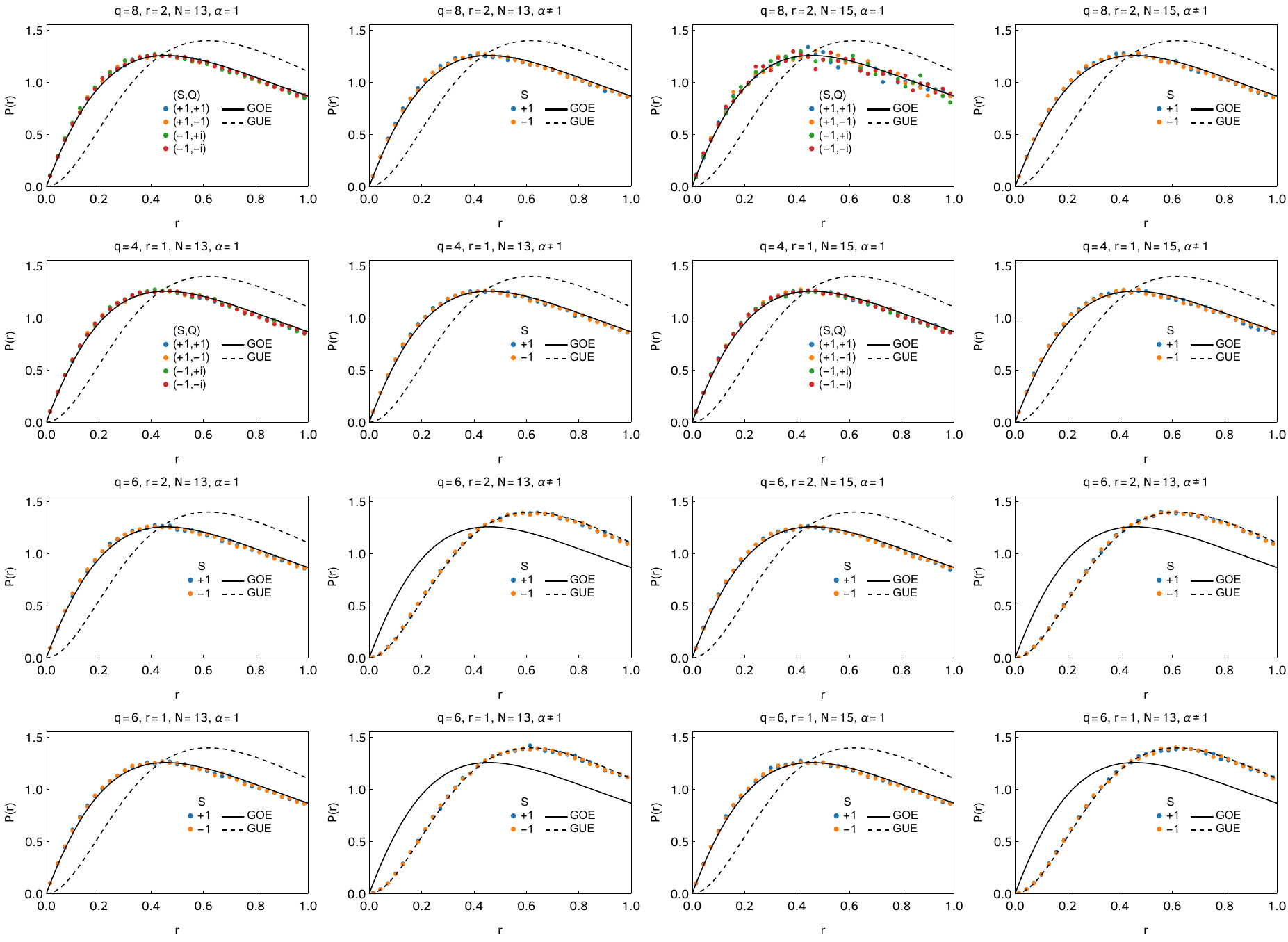}
	\caption{Same as Fig.~\ref{fig:SM_LevelStats_EvenN}, but for odd $N$. We find excellent agreement with the RMT predictions for all cases, see Tables~\ref{tab:classification_odd_even} and \ref{tab:classification_odd_odd}.
	}
	\label{fig:SM_LevelStats_OddN}
\end{figure}

To probe the bulk local level correlations of the Hamiltonian~(\ref{eq:hamdef}) for different values of the parameters $N$, $q$, $r$, and $\alpha$, and confirm the classification put forward in Tables~\ref{tab:classification_even_even_even}--\ref{tab:classification_even_odd_odd}, \ref{tab:classification_odd_even}, and \ref{tab:classification_odd_odd}, we compute the spacing ratio distribution, $P(r)$, where~\cite{oganesyan2007}
\begin{equation}
r_i=\min\left(\frac{E_{i+1}-E_i}{E_i-E_{i-1}},\frac{E_i-E_{i-1}}{E_{i+1}-E_i}\right),
\end{equation}
for the ordered eigenvalues $E_i$ of a block of the Hamiltonian. This observable is complementary to the unfolded spacing distribution considered in the main text and, conveniently, does not require unfolding of the spectrum.

We obtained $P(r)$ numerically from exact diagonalization, performing an ensemble average over the disordered couplings $J_{i_1\cdots i_q}$, collecting around $2^{22}$ eigenvalues for each set of parameters. To avoid boundary effects we discarded the first and last $1/16$ of the eigenvalues of each block.
We compare the numerical results with the predictions of RMT in the form of the Wigner-like surmise~\cite{atas2016}
\begin{equation}
\label{eq:surmise}
P(r)=\frac{2}{Z_\beta}\frac{(r+r^2)^\beta}{(1+r+r^2)^{3\beta/2}},
\end{equation}
with $\beta=1$ and $Z_1=8/27$ for the GOE, and $\beta=2$ and $Z_2=4\pi/81\sqrt{3}$ for the GUE. The results are depicted in Figs.~\ref{fig:SM_LevelStats_EvenN} and \ref{fig:SM_LevelStats_OddN} for even $N$ (compare with Tables~\ref{tab:classification_even_even_even}--\ref{tab:classification_even_odd_odd}) and odd $N$ (Tables~\ref{tab:classification_odd_even} and \ref{tab:classification_odd_odd}), respectively. We find excellent agreement in all cases.

\subsection{Hard-edge correlations: distribution of the smallest eigenvalue}\label{sec:2.2}

To distinguish universality classes BDI, CI, C, and D, which have special spectral features near $E_0=0$, from the bulk Wigner-Dyson classes A and AI, we computed the microscopic spectral density near $E_0$ in the main text. Here, we present an alternative confirmation of the symmetry classification of Table~\ref{tab:classification_even_odd_odd} in terms of the distribution of the eigenvalue closest to $E_0$ in units of its average value, denoted $E_1>0$, which again does not require unfolding. We recall that these classes arise only for odd $q/2$ and odd $r$.

\begin{figure}[t]
	\centering
	\includegraphics[width=\textwidth]{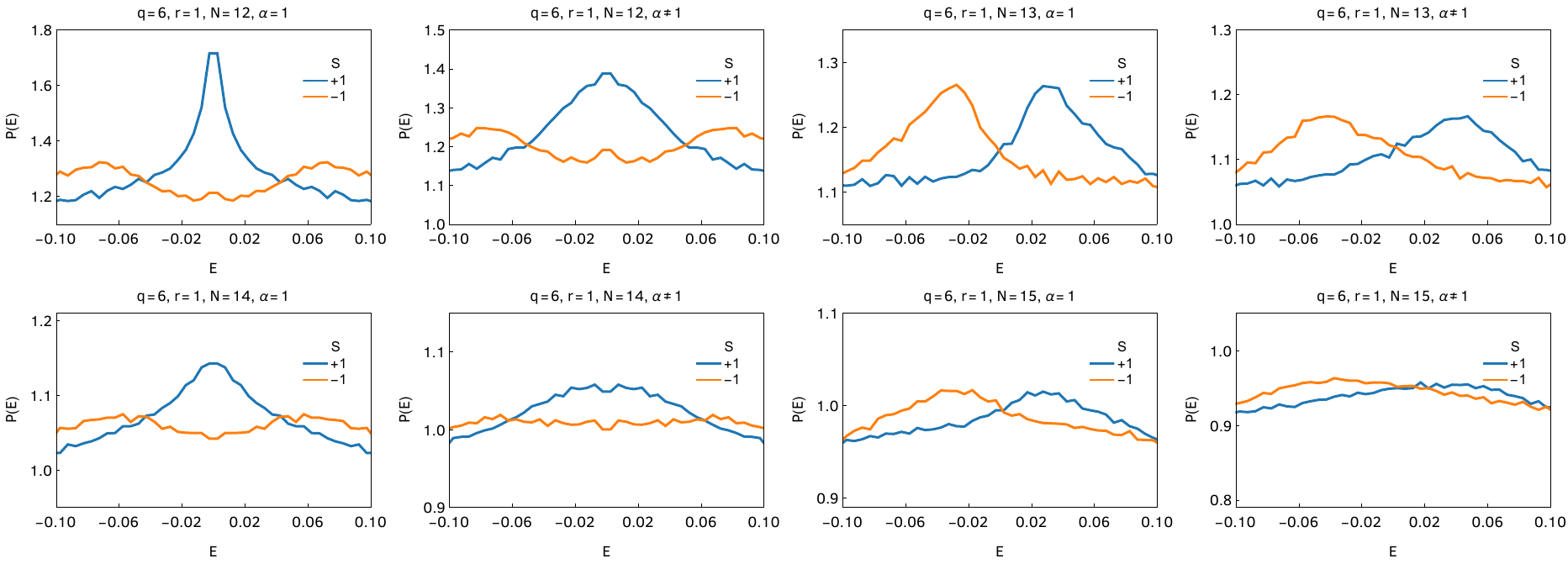}
	\caption{Spectral density around of $E_0=0$, for $q=6$, $r=2$, and $N=12$--$15$. For all cases $\lambda=0.05$ and we take $\alpha=1.1$ whenever $\alpha\neq1$ is indicated. The colored curves represent the numerical results obtained from exact diagonalization for the different blocks of the Hamiltonian, labeled by the total parity $S=\pm1$. The spectral density is symmetric around $E_0$, but the pair $(E,-E)$ only belongs to the same block if $N$ is even, regardless of the value of $\alpha$.
	}
	\label{fig:SM_LevelDen}
	\includegraphics[width=\textwidth]{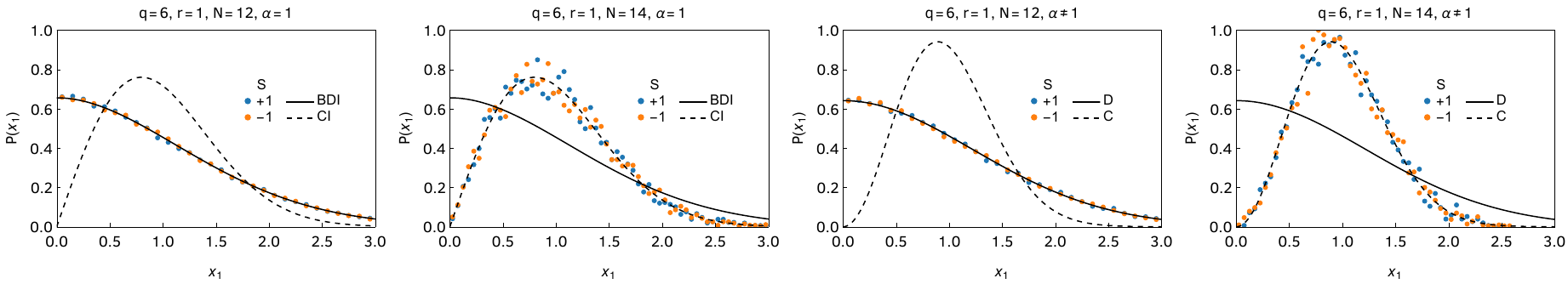}
	\caption{Distribution of the eigenvalue closest to $E_0=0$ normalized to unit mean, $x_1=E_1/\av{E_1}$, for $q=6$, $r=2$, and $N=12$ and $14$. For all cases $\lambda=0.05$ and we take $\alpha=1.1$ whenever $\alpha\neq1$ is indicated. The colored dots represent the numerical results obtained from exact diagonalization for the different blocks of the Hamiltonian, while the black (full and dashed) curves give exact RMT results for classes BDI and CI, Eqs.~(\ref{eq:emin_BDI}) and (\ref{eq:emin_CI}), (for $\alpha=1$) or the RMT surmises for classes C and D, Eqs.~(\ref{eq:emin_C}) and (\ref{eq:emin_D}), ($\alpha\neq1$). We find excellent agreement with the RMT predictions for all cases, see Table~\ref{tab:classification_even_odd_odd}.
	}
	\label{fig:SM_Emin}
\end{figure}

First, we check that for odd $N$, the PHS operator $C$ does not act within a single block of $H$ and, hence, does not define a symmetry class (in this case, BDI, CI, C, or D). To do so, we show, in Fig.~\ref{fig:SM_LevelDen}, the spectral density close to $E_0$. We see that it is symmetric around $E_0$, which signals the presence of PHS. However, only for even $N$, do pairs $(E_i,-E_i)$ belong to the same block of the Hamiltonian (labeled by $S=\pm1$) and $C$ acts within a single block, while for odd $N$, the pairs belong to blocks with opposite parities and $C$ connects different blocks, as discussed before. The reason that the
spectral density for $S=1$ is different from the spectral density for $S=-1$ is that the value of the coupling $\lambda $ is relatively large so that the structure of the spectrum of
$
i\sum_{k=1}^N \psi_k^L \psi_k^R
$
remains visible in the level density of the total Hamiltonian~\cite{garcia2019}.

Next, we compute the probability density function $P(x_1)$ of the eigenvalue $E_1$ normalized to unit mean, $x_1=E_1/\av{E_1}$. We computed them numerically from exact diagonalization by sampling around $10^4$ realizations of $H$ for $q=6$, $r=1$, and $N=12$ and $N=14$. The intersite coupling is taken small with strength $\lambda=0.05$, while $\alpha=1.1$ whenever $\alpha\neq1$ is indicated. For the RMT ensembles belonging to classes BDI, CI, C, and D, these distributions are known exactly. They are given by~\cite{forrester1993NPB}
\begin{equation}
\label{eq:emin_BDI}
P_\mathrm{BDI}(x_1)=a(2+bx_1)\exp{-\frac{b^2x_1^2}{8}-\frac{bx_1}{2}},
\qquad
a=\frac{1}{2}\sqrt{\frac{\pi e}{2}}\ \mathrm{erfc}(1/\sqrt{2}),
\qquad
b=\sqrt{2\pi e}\ \mathrm{erfc}(1/\sqrt{2}),
\end{equation}
for class BDI and by~\cite{wilke1998PRD,nishigaki1998PRD,damgaard2001PRD}
\begin{equation}
\label{eq:emin_CI}
P_\mathrm{CI}(x_1)=a\ b\ x_1 \exp{-2b^2x_1^2},
\qquad
a=\sqrt{2\pi},
\qquad
b=\sqrt{\pi/8},
\end{equation}
for class CI. The constant $a$ fixes the normalization of $P(x_1)$, while $b$ fixes the mean of $x_1$ to be one. The exact expressions for classes C and D are obtained from the derivative of a Fredholm determinant~\cite{sun2020}, but simple closed-form expressions can be derived from a Wigner-like surmise~\cite{akemann2009PRE} for $2\times2$ matrices. These are given by~\cite{sun2020}
\begin{equation}
\label{eq:emin_C}
\begin{split}
&P_\mathrm{C}(x_1)=a b^2 x_1^2 \exp{-2 b^2 x_1^2}
\left[30 b x_1 - 4 b^3 x_1^3
+ \sqrt{\pi} \exp{b ^2 x_1^2} \mathrm{erfc}(b x_1)
\left(15 - 12 b^2 x_1^2 + 4 b^4 x_1^4\right)\right],
\\
&a=\frac{10-5\sqrt{2}}{3\pi^{3/2}},
\qquad
b=\frac{10-5\sqrt{2}}{2\sqrt{\pi}},
\end{split}
\end{equation}
for class C and by
\begin{equation}
\label{eq:emin_D}
\begin{split}
&P_\mathrm{D}(x_1)=a b^2\exp{-2 b^2 x_1^2}
\left[6 b x_1 - 4 b^3 x_1^3
+ \sqrt{\pi} \exp{b ^2 x_1^2} \mathrm{erfc}(b x_1)
\left(3 - 4 b^2 x_1^2 + 4 b^4 x_1^4\right)\right],
\\
&a=\frac{7-4\sqrt{2}}{2\pi^{3/2}},
\qquad
b=\frac{7-4\sqrt{2}}{2\sqrt{\pi}},
\end{split}
\end{equation}
for class D.
The comparison of the numerical and analytical results is given in Fig.~\ref{fig:SM_Emin}, with perfect agreement in all cases.

\section{Numerical solution of the large-$N$ Schwinger-Dyson equations: gap and free energy}

In this appendix, we discuss technical details of the numerical solutions of the SD equations of two coupled SYK models.
We will show the existence of a first-order phase transition
and discuss the extraction of the gap from the long-time behavior of the Green's function.

The finite-temperature partition function $Z$ of the Hamiltonian Eq.~(\ref{eq:hami})
is evaluated using the path integral formalism for fermion fields. Following the standard procedure~\cite{maldacena2016,maldacena2018},
the path integral is evaluated by expressing the fermion bilinears in terms of
\be
G^{ab}(\tau_1,\tau_2)= \frac{1}{N}\sum_{i=1}^{N}\psi^a_i(\tau_1)\psi^b_i(\tau_2)
\ee
 using
the Lagrange multipliers $\Sigma(\tau_1,\tau_2)$ 
($a,b =L,R$). We obtain the action $S[G,\Sigma]$ given by (note that $S$ is the normalized action divided by a factor of $N$)
\begin{equation}
\langle Z\rangle \sim \int DG D\Sigma e^{-NS[G,\Sigma]},
\end{equation}
with
\begin{equation}
\begin{aligned}
  {S}=& -\frac{1}{2}\int d\tau \log \det \(\partial_{\tau}  -\Sigma^{ab}\)^2
 +\frac 12 \int d\tau_1 d\tau_2 \left [  \Sigma^{ab}(\tau_1,\tau_2)G^{ab}(\tau_1,\tau_2)
-s_{ab}\frac{2^{q-1}}{q^2}\(G^{ab}(\tau_1,\tau_2)\)^q\right ]\\
    &-\frac{i^r\lambda}{2 r}\int d\tau\left [  \(G^{RL}(\tau.\tau)\)^r  +\(- G^{LR}(\tau,\tau)\)^r\right ].
\end{aligned}
\label{eq:S}
\end{equation}
Here, $s_{LL}=s_{RR}=1$ and $s_{LR}=s_{RL}=(-1)^{q/2}$.

The saddle point equation are simplified by using translational invariance,
\begin{equation}
  G^{ab}(\tau)=G^{ab}(\tau_1-\tau_2)
\end{equation}
as well as other symmetries of $G^{ab}(\tau_1,\tau_2)$.
This results in the Schwinger-Dyson (SD) equations,
\begin{equation}\label{eq:SD}
\begin{aligned}
&G^{LL}(\omega)=\frac{-i\omega -\Sigma^{LL}(\omega)}{D(\omega)},
\qquad
G^{LR}(\omega) =\frac{\Sigma^{LR}(\omega)}{D(\omega)},
\qquad
D(\omega) = (i\omega +\Sigma^{LL}(\omega))^2 +(\Sigma^{LR}(\omega))^2, \\
&\Sigma^{LL}(\tau)=\frac{2^{q-1}}{q}G^{LL}(\tau)^{q-1},
\qquad
\Sigma^{LR}(\tau)=(-1)^{\frac{q}{2}} \frac{2^{q-1}}{q}G^{LR}(\tau)^{q-1} +i^r \lambda G^{LR}(\tau=0)^{r-1}\delta(\tau).
\end{aligned}
\end{equation}
\begin{figure}[t!]
	\centering
	\subfigure[]{\includegraphics[width=8cm]{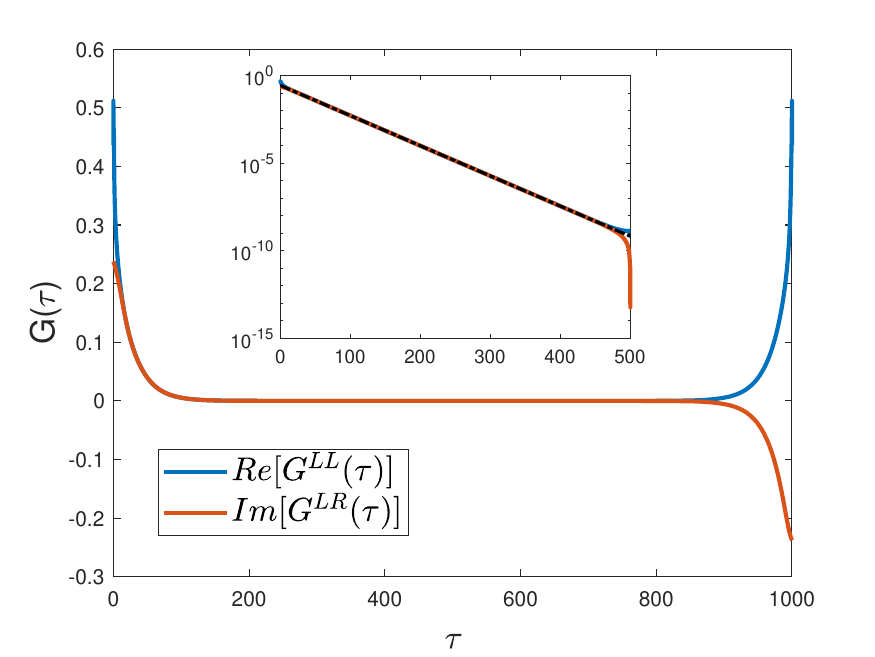}}
	\caption{The Green's function $G^{ab}(\tau)$ in the wormhole phase, obtained by solving the SD equations~(\ref{eq:SD}) of two coupled
          SYK models
          with $q=6,\; r=2, \; T=10^{-3},\; \lambda=0.08$ and $ M=2^{20}$.
          The Green's functions $|G^{LL}|$ and $|G^{LR}|$ are close except when
          $\tau$ is near zero or $\beta$. The black dotted curve in the inset represents
          a first-order polynomial fit to the exponential decaying part
          of $\log |G^{LL}(\tau)|$ and $\log|G^{LR}(\tau)|$.}
        \label{fig:GreenSigma}
\end{figure}
These equations are solved numerical by discretizing the $\tau$ and $\omega$ variables 
according to
\begin{equation}\label{eq:disc}
\begin{aligned}
&\omega_n =\frac{2\pi(n+\frac{1}{2})}{\beta}  \qquad n =-\frac{M}{2},-\frac{M}{2}+1\dots,\frac{M}{2}-1 \\
&\tau_m=\frac{(m+\frac 12) \beta}{M} \qquad  m =0, 1,\dots ,M-1,
\end{aligned}
\end{equation}
with $\beta$ the inverse temperature and $M$ the number of discretization points.
The Fourier transform is calculated by a Fast Fourier Transformation (FFT) algorithm,
and the
$i^r\lambda [G^{LR}(0)\delta(\tau)]^{r-1}$ term is included after Fourier transformation.
The $\frac 12$ increment in $\tau_m$ eliminates the Gibbs effect in the numerical Fourier transforms. As a result, in Fig.~\ref{fig:GreenSigma}, the fluctuations of $G^{ab}(\tau)$  near $\tau=0 $
and $\beta$ are suppressed. To increase the robustness of the algorithm, we impose the symmetries
of the Green's function after each iteration step. 

In Fig.~\ref{fig:GreenSigma} we show the Green's functions $G^{LL}(\tau)$ and $G^{LR}(\tau)$
for $q=6$ and $r=2$ obtained by solving the SD equations. In the inset of this figure, we note
that both Green's functions decay exponentially over a wide range of time, 
$G^{ab}(\tau)\propto \exp{-E_g \tau}$. The gap $E_g$ is obtained by linear fit to this part
of these curve. Since $G^{LL}(\tau) $ and $G^{LR}(\tau) $ almost coincide in the exponential
decaying region, both Green's functions give the same gap. 

The free energy, $F= S[G,\Sigma]/\beta$, is computed by substituting the saddle-point results for
$G^{ab}(\tau)$ and $\Sigma^{ab}(\omega)$ into Eq.~(\ref{eq:S}) [with discretizations $\tau_m$
 and $\omega_n$ given in Eq.~(\ref{eq:disc})] resulting in,
\begin{equation}\label{eq:Free}
\begin{aligned}
F =   -\frac{1}{\beta}\Biggl[&\ln2 +\frac{1}{2}\sum_{\omega_n}\ln \biggl(\frac{(i\omega_n+\Sigma^{LL}(\omega_n))^2+(\Sigma^{LR}(\omega_n))^2}{-\omega_n^2}\biggr) +\sum_{\omega_n} \Bigl(\,\Sigma^{LL}(\omega_n)G^{LL}(\omega_n)-\Sigma^{LR}(\omega_n)G^{LR}(\omega_n)\Bigr)\Biggr]\\
&- \frac{2^{q-1}}{q^2} \int_0^\beta
\biggl(\,(G^{LL}(\tau))^q +(-1)^{q/2}(G^{LR}(\tau))^q\biggr)d\tau- \frac{i^r\lambda}{r}(G^{LR}(\tau=0))^r.
\end{aligned}
\end{equation}

\begin{figure}[t!]
	\centering
	\subfigure[]{\includegraphics[width=8cm]{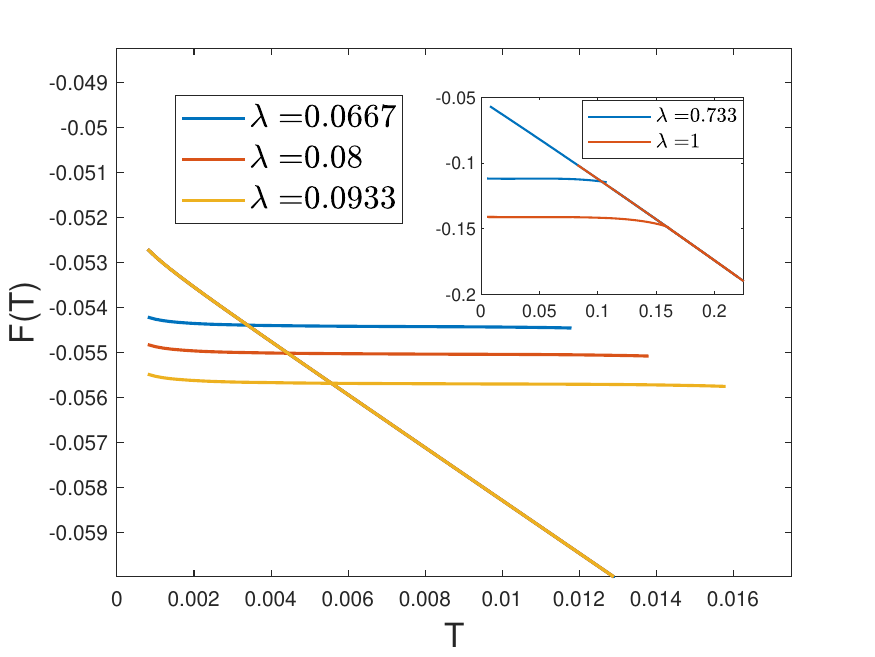}}
	\subfigure[]{\includegraphics[width=8cm]{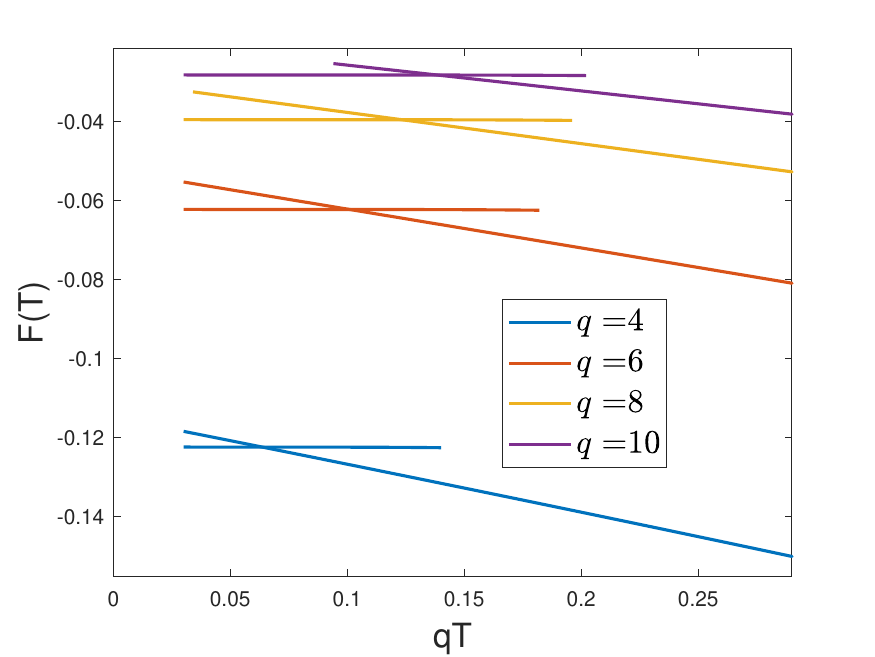}}
	\caption{Free energy as a function of temperature $T$ for
two coupled SYK models with $r=2$ and different values of $q$ and the coupling strength $\lambda$. The number of discretization points is equal to $M=2^{16}$. In the left panel (a), we compare different values of $\lambda$, for $q=6$ and $r=2$. In the wormhole phase, the free energy is independent of the temperature and decreases as $\lambda$ increases,
          while in the high-temperature phase, it is independent of $\lambda$ and depends linearly on temperature. In the inset, we show that the first-order transition occurs only
          for sufficiently weak coupling, $\lambda < \lambda_c \approx 0.73$. For larger coupling the transition becomes a crossover. In the right panel (b), we compare different
          values of $q$, with $\tilde{\lambda}\equiv q/2^{r-1}\lambda = 0.6$. The transition temperature is not very sensitive to $qT$. The free energy becomes smaller as we increase $q$.}\label{fig:freeSD}
\end{figure}

In order to find out the temperature dependence of the free energy, we choose the ``seed'' method for iteration, namely, we begin with the SD solution $G^{ab}(\tau)$ at temperature $T_i$,
and use it as the initial guess for the solution of the SD equations at
$T_{i+1}=T_i+\Delta T$, and, after having converged on a solution of the SD equations
for $T_{i+1}$, we use this solution as the starting point for $T_{i+2}=T_i+2\Delta T$, and so on.
The temperature step $\Delta T$ should be small to find continuous branches of $F(T)$.
In practice, we first increase the temperature by choosing $\Delta T>0$, and after reaching a sufficiently high temperature, we decrease the temperature with $\Delta T<0$.
In Fig.~\ref{fig:freeSD}, we find the intersection between two phases for different parameters ($q,r,\lambda$). If there exist several solutions of the SD equations for a fixed temperature, we choose the one with the lower free energy.
As a consequence of the crossing of two branches of the free energy at a certain temperature, corresponding to two different solutions of the SD equations, the system undergoes a first-order phase transition~\cite{maldacena2018}
between the traversable wormhole phase characterized by a flat, almost temperature independent,
free energy in the low-temperature limit, and the black hole phase at higher temperature,
for which the free energy decreases close to linearly.
In Fig.~\ref{fig:freeSD} we show the free energy as a function of
the temperature for $r=2$ and various values of $q$ and $\lambda$. From the inset
of Fig.~\ref{fig:freeSD}~(a) we see that when $\lambda$ is large enough, the first-order transition becomes a crossover, as in the $r=1$ case~\cite{maldacena2018,garcia2019}.
These results confirm that the wormhole phase requires low temperature and weak intersite coupling.

\end{document}